\documentclass[preprint]{aastex631}
\usepackage[T1]{fontenc}
\usepackage{appendix}
\usepackage[utf8]{inputenc}
\usepackage{afterpage}
\usepackage{placeins}
\usepackage{gensymb}
\pdfoutput=1

\begin{document}

\title{Estimate of water and hydroxyl abundance on asteroid (16) Psyche from JWST data}

\author[0000-0002-0786-7307]{Stephanie G. Jarmak}
\affiliation{Southwest Research Institute \\
San Antonio, TX, USA}
\affiliation{Harvard \& Smithsonian | Center for Astrophysics \\
Cambridge, MA, USA}

\author[0000-0002-1559-5954]{Tracy M. Becker}
\affiliation{Southwest Research Institute \\
San Antonio, TX, USA}

\author[0000-0001-6567-627X]{Charles E. Woodward}
\affiliation{Minnesota Institute for Astrophysics, University of Minnesota Twin Cities \\ 
Twin Cities, MN, USA 55455}

\author[0000-0001-8248-8991]{Casey I. Honniball}
\affiliation{NASA Goddard Space Flight Center \\ 
Greenbelt, MD, USA}

\author[0000-0002-9939-9976]{Andrew S. Rivkin}
\affiliation{The John Hopkins University Applied Physics Laboratory \\
Laurel, MD, USA}

\author[0000-0003-3356-1368]{Margaret M. McAdam} 
\affiliation{NASA AMES Research Center \\
Mountainview, CA, USA}

\author[0000-0003-4980-1135]{Zoe A. Landsman} 
\affiliation{University of Central Florida \\
Orlando, FL, USA}

\author[0000-0001-6294-4523]{Saverio Cambioni} 
\affiliation{Massachusetts Institute of Technology \\
Cambridge, MA, USA}

\author[0000-0002-0717-0462]{Thomas G. Müller} 
\affiliation{Max-Planck Institute for Extraterresetrial Physics \\
Garching, Germany}

\author[0000-0003-4942-2741]{Driss Takir} 
\affiliation{NASA Johnson Space Center \\
Houston, TX, USA}

\author[0000-0001-9470-150X]{Kurt D. Retherford} 
\affiliation{Southwest Research Institute \\
San Antonio, TX, USA}

\author[0000-0002-1706-6255]{Anicia Arredondo} 
\affiliation{Southwest Research Institute \\
San Antonio, TX, USA}

\author[0000-0003-4008-1098]{Linda T. Elkins-Tanton} 
\affiliation{Arizona State University \\
Tempe, AZ, USA}



\begin{abstract}

Our understanding of Solar System evolution is closely tied to interpretations of asteroid composition, particularly the M-class asteroids. These asteroids were initially thought to be the exposed cores of differentiated planetesimals, a hypothesis based on their spectral similarity to iron meteorites. However, recent astronomical observations have revealed hydration on their surface through the detection of 3-$\mu$m absorption features associated with OH and potentially H\textsubscript{2}O. We present evidence of hydration due mainly to OH on asteroid (16) Psyche, the largest M-class asteroid, using data from the James Webb Space Telescope (JWST) spanning 1.1 - 6.63 $\mu$m. Our observations include two detections of the full 3-$\mu$m feature associated with OH and H\textsubscript{2}O resembling those found in CY-, CH-, and CB-type carbonaceous chondrites, and no 6-$\mu$m feature uniquely associated with H\textsubscript{2}O across two observations. We observe 3-$\mu$m depths of between 4.3 and 6\% across two observations, values consistent with hydrogen abundance estimates on other airless bodies of 250 - 400 ppm. We place an upper limit of 39 ppm on the water abundance from the standard deviation around the 6-$\mu$m feature region. The presence of hydrated minerals suggests a complex history for Psyche. Exogenous sources of OH-bearing minerals could come from hydrated impactors. Endogenous OH-bearing minerals would indicate a composition more similar to E-or P-class asteroids. If the hydration is endogenous, it supports the theory that Psyche originated beyond the snow line and later migrated to the outer main belt.

\end{abstract}
\keywords{Asteroids (72); Infrared spectroscopy (2285); Surface composition (2115)}


\NewPageAfterKeywords

\section{Introduction} \label{sec:intro}

(16) Psyche (hereafter Psyche) is the largest M-class asteroid in the Tholen taxonomy \citep{1984PhDT.........3T}. This class of asteroids has long been believed to include the remnant metallic cores of large, differentiated asteroids exposed through a series of collisions that stripped off their mantles e.g., \cite{1973Icar...19..507C}, \cite{1990JGR....95..281C}, and \cite{2015IAUGA..2256152S}. They are also thought to include the parent bodies of the iron meteorites \citep{1989aste.conf..921B}. Observations supporting a metallic composition for Psyche are its high radar albedo (0.34 ± 0.08; \cite{2021PSJ.....2..125S}), bulk density potentially higher than expected from a solar abundance of iron and rocky forming elements (4.0 ± 0.2 g/cm\textsuperscript{3 }; \cite{2021PSJ.....2..125S}), spectral matches to pure iron and iron meteorites e.g., \cite{2010Icar..210..655F}, \cite{2020PSJ.....1...53B}, metal content of no less than 20\% as inferred from millimeter wavelength data \citep{2021PSJ.....2..149D}, and mid-infrared (MIR) spectroscopy from \cite{2024PSJ.....5...33A} indicating a lack of strong spectral features consistent with a metal/oxide surface. The evidence that Psyche is metal-rich and could be the exposed iron core of a differentiated asteroid has generated significant community interest culminating in the \textit{Psyche} mission that launched October 2023 with an expected arrival date of August 2029 e.g., \cite{2023E&SS...1002694D}.

However, there is also evidence that challenges this view. M-class asteroids, while consistent with iron meteorites due to their lack of features in the vis-NIR wavelength range used in the Tholen taxonomy, are also consistent with enstatite chondrites that are similarly featureless in these wavelengths. Additionally, Psyche’s density measurements have varied widely over the years, and recent estimates suggest a density of 3.88 ± 0.25 g/cm\textsuperscript{3} \citep{2020A&A...633A..46S}, which if towards the lower end, would be similar to that of Vesta, a non-metallic asteroid. This raises doubts about Psyche being predominantly metallic (> 50\%), as achieving such a low density would require an unrealistically high macroporosity, unlikely for an object of Psyche’s size without evidence of a massive asteroid family that would result from its breakup and reaccumulation \citep{1999Icar..137..140D}, \citep{2018MNRAS.475.3419A}. Additionally, spectral features attributed to silicate minerals have also been detected on Psyche and other M-class asteroids e.g., \cite{2011M&PS...46.1910H}, \cite{2017AJ....153...29S}, \cite{2018Icar..304...58L}, and references therein. Thermal inertia values derived from disk-integrated mid-infrared data (125 ± 40 J s\textsuperscript{1/2} K\textsuperscript{-1}m\textsuperscript{-2} \citep{2013Icar..226..419M}; 5 - 25 J s\textsuperscript{1/2} K\textsuperscript{-1}m\textsuperscript{-2 }\citep{2018Icar..304...58L}; 20 - 80 J s\textsuperscript{1/2} K\textsuperscript{-1}m\textsuperscript{-2 }\citep{2022A&A...659A..38R}) are at the lower end of what is expected for a powdered metal regolith (e.g, < 450 J s\textsuperscript{1/2} K\textsuperscript{-1}m\textsuperscript{-2}, \citep{2013LPI....44.1018C}) and suggest the presence of at least some silicates. \cite{2022JGRE..12707091C} found that the thermal inertia across Psyche’s surface was 25 - 600 s\textsuperscript{1/2} K\textsuperscript{-1}m\textsuperscript{-2 }based on spatially-resolved millimeter wavelength data, with most of Psyche’s surface thermal inertia falling in the 150 - 300 J s\textsuperscript{1/2} K\textsuperscript{-1}m\textsuperscript{-2  }range and dielectric constant (proxy for metal content) in the 15-25 range, also consistent with a heterogeneous surface in terms of metal content.

Near-infrared (NIR) surveys of M-class asteroids (\cite{1990Icar...88..172J}, \cite{1995Icar..117...90R}, \cite{2000Icar..145..351R}, \cite{2015Icar..252..186L}, \cite{2019PASJ...71....1U}) detected 3-$\mu$m spectral absorption features attributed to hydrated silicates on $\sim$35\% of known M-class asteroids. Interestingly, Psyche was one of the only two large (D > 65 km) M-class asteroids in the \cite{2000Icar..145..351R} study for which they did not detect a 3-$\mu$m band, though the uncertainties on these spectrophotometric measurements precluded the detection of weak absorption features. Later high-resolution spectral observations by \cite{2017AJ....153...31T} reported the detection of a shallow 3-$\mu$m feature on Psyche with apparent rotational variability. This feature is difficult to detect and impossible to characterize fully from Earth-based telescopes due to atmospheric water absorptions blocking a part of the 3-$\mu$m feature. The AKARI spacecraft did not detect a 3-$\mu$m feature but also concluded the shallow feature reported by \cite{2017AJ....153...31T} would be below their detection limit \citep{2019PASJ...71....1U}. The detection of a 3-$\mu$m feature implies the presence of hydroxyl (OH) or water-bearing minerals, depending on the precise center and shape of the band \citep{2012Icar..219..641T}, \citep{2018Icar..304...74R}. A more unambiguous detection of H\textsubscript{2}O would come from detecting the H\textsubscript{2}O 6-$\mu$m fundamental bending mode emission feature. This 6-$\mu$m feature was used to detect widespread molecular water on the Moon \citep{2021NatAs...5..121H} and recently on asteroids \citep{2024PSJ.....5...37A}. 

The presence and strength of either of these features might be linked to several possible formation and evolutionary history scenarios for the Solar System. Assessing the abundance of hydration products present on Psyche provides important context for the amount of volatiles that the largest planetesimals could retain during their formation and differentiation in the early solar system and thus whether impacts of differentiated planetesimals onto the early Earth should be considered an important source of water. Hydrous materials, whether water- or OH-bearing minerals or the ices themselves, may be remnants of planetesimal formation or they could be later and even recent additions from impacts. \cite{2022PSJ.....3..196B} found that water ice is not stable at any latitude on Psyche, and therefore any detected water is most likely in the form of water-bearing minerals. The two competing hypotheses for Psyche’s hydrated surface are either exogenous sources via C-complex asteroids in the neighborhood of Psyche e.g., \citep{2018MNRAS.475.3419A}, or that Psyche formed in the outer solar system and retained hydrated materials upon differentiation e.g., \cite{2022SSRv..218...17E}. The recently-launched\textit{ Psyche }mission will obtain elemental, space physics, and multispectral imaging data with high spatial resolution at the asteroid, but does not have spectral capabilities beyond 1.1 $\mu$m. Complementary measurements from other facilities are thus necessary to comprehensively understand the abundance and nature of hydrated minerals on Psyche's surface.

\section{Methods} \label{sec:style}

\begin{figure}[!h]
    \centering
    \includegraphics[width=1\linewidth]{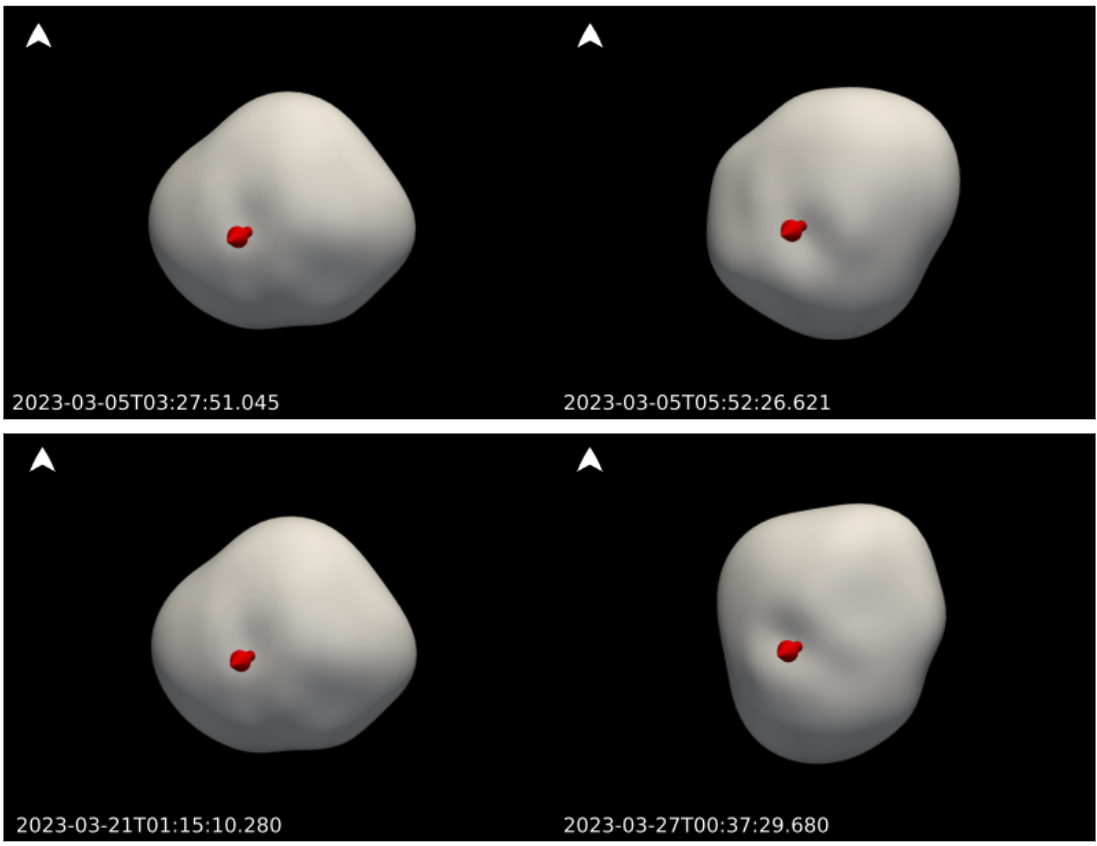}
    \caption{Orientation of Psyche at the time of observations (Top) NIRSpec observations (Bottom) MIRI observations. Each observation is nearly pole-on (aspect angle 24°) viewing the north pole. These figures were produced in Pyvista \citep{2019JOSS....4.1450S} using the shape model from \cite{2021PSJ.....2..125S}.}
    \label{fig:PsycheOrientation_Fig1}
\end{figure}

We observed Psyche with the James Webb Space Telescope (JWST) using the Near Infrared Spectrograph (NIRSpec) and Mid-Infrared Instrument (MIRI) instruments to fully characterize the 3-$\mu$m absorption feature associated with OH/H\textsubscript{2}O, and to identify whether the 6-$\mu$m emission feature associated with H\textsubscript{2}O is present. We obtained two sets of observations for each instrument to assess the variability of these features with the NIRSpec observations each on March 5 2023, and the MIRI observations on March 20 and 27, 2023. During these observations, Psyche was nearly pole-on (aspect angle 24°) with observations covering the north pole region (Figure \ref{fig:PsycheOrientation_Fig1}).

\FloatBarrier
\FloatBarrier

\begin{figure}[!h]
    \centering
    \includegraphics[width=1\linewidth]{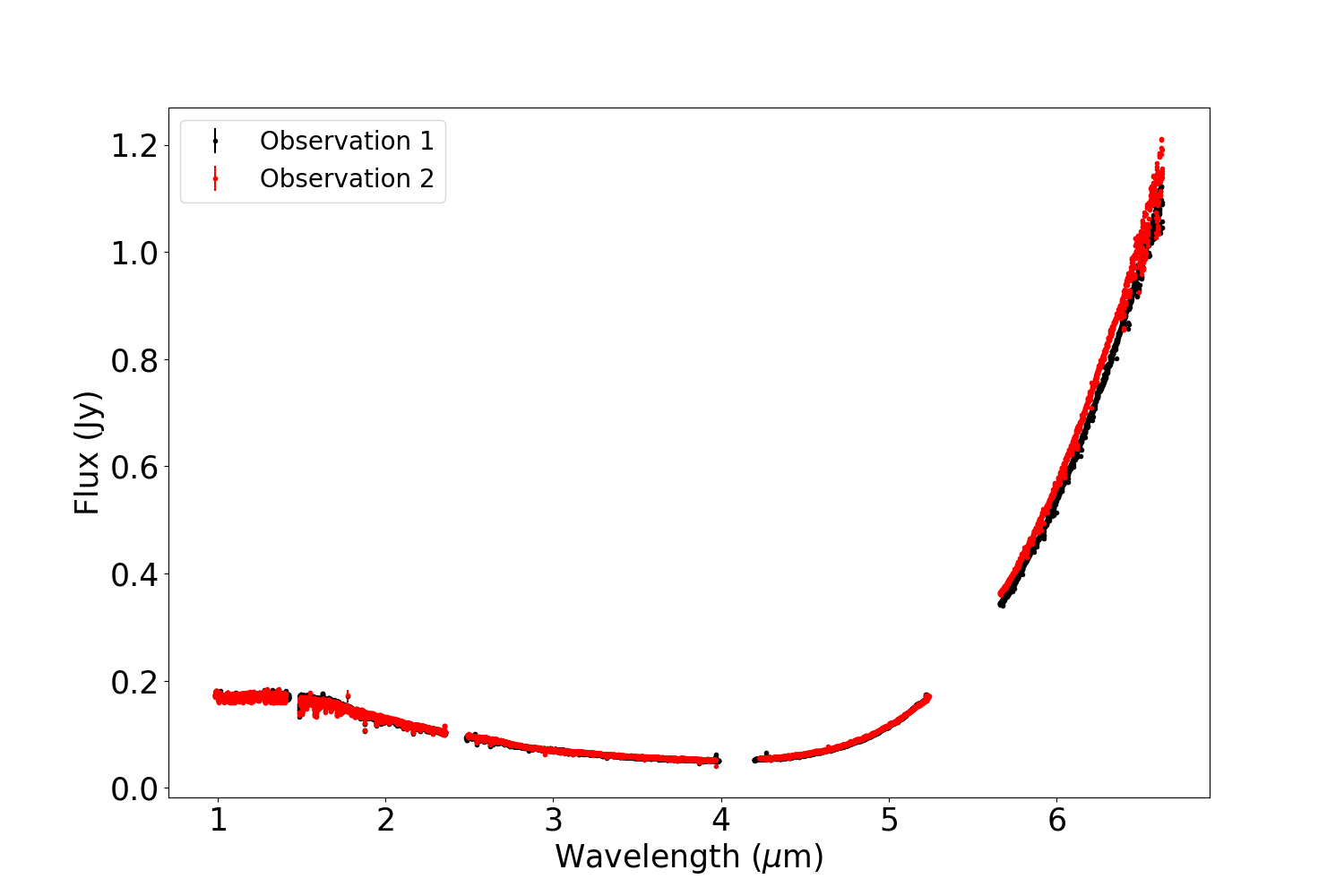}
    \caption{Spectral energy distribution of Psyche obtained from NIRSpec and MIRI instruments prior to removal of the solar spectrum and thermal emission to create a reflectance spectrum. Average resolution is $\sim$0.001 $\mu$m for the NIRSpec observations and $\sim$0.002 $\mu$m for the MIRI observations, corresponding to a factor of $\sim$25x (AKARI) and $\sim$60x (Spitzer) increased resolution compared to previous space telescope infrared observations of Psyche. For each instrument, observation 1 is in black and observation 2 is in red. }
    \label{fig:PsycheFlux_Fig2}
\end{figure}

We show the spectrum acquired with the NIRSpec and MIRI instruments, featuring scattered sunlight and thermal emission from the asteroid, in Figure \ref{fig:PsycheFlux_Fig2}. 

\FloatBarrier
\FloatBarrier

\subsection{Observations and Data Reduction} \label{subsec:Observations and data reduction}

We used JWST’s NIRSpec and MIRI instruments to observe Psyche (program ID 1731) in March 2023. The NIRSpec Integral Field Unit observations produced two data sets with the first observation beginning March 5 2023 02:21:33 UT and the second observation beginning March 5 2023 03:15:10 UT using the G140H/F100LP, G235H/F170LP, and G395H/F290LP modes, each with a total effective exposure time of 128.84 s. The MIRI Integral Field Unit observations also produced two data sets, with the first observation starting March 20 2023 22:51:59 UT and the second observation starting March 26 20:39:46 UT using all four spectral channels of the MRS (and corresponding sub-bands) with an exposure time of 632.71 s. The target saturated in Channels 2 through 4 and improving the correspondingly degraded data quality to carry out spectral feature analysis requires effort beyond the scope of our program goals. The proposed MIRI observations included Channel 1 (B = MED) data only and given that a 6-$\mu$m feature would be entirely detectable within Channel 1 (expected centers are 6.04 - 6.12 $\mu$m with widths of 0.1 - 0.55 $\mu$m, see Figure 2 \citep{2022GeoRL..4997786H}) we therefore focus only on analysis of Channel 1 observations.  
 
Uncalibrated (level-0) PID 1731 data products (\_ucal files) were retrieved from the
Mikulski Archive for Space Telescopes and reprocessed with JWST Science Calibration
Pipeline calibration versions (CAL\_VER) v1.11.3 and v1.14.3 for the NIRSpec IFU and
MIRI MRS IFU data respectively. The MRS data were locally processed with Calibration
Reference Data System (CRDS) file CRDS\_CTX jwst\_1241.pmap, while the NIRSpec
data used CRDS\_CTX jwst\_1106.pmap.
In Stage 2 of the pipeline (Spec2Pipeline), the NIRSpec rate files where de-stripped
(suppressing the vertical stripes in the frames from 1/f noise) and background
subtraction was achieved using rate\_combinedbackground files generated from the
background data model using the four off-source background observations for each
grating. The final NIRSpec IFU spectral cubes (\_s3d file) were then generated in
Stage3 of the pipeline (Spec3Pipeline) using the \_cal files generated in Stage2 sorted
into proper association files (\_asn.json) for NRS1 and NRS2, where the
outlier\_detection and master\_background steps were both skipped. The MRS IFU data
were processed in a similar fashion, where generation of the rate files from the level\_0 in
the Detector1Pipeline stage invoked a jump.three\_group\_rejection\_threshold = 100
(useful for very bright targets and short ramp times) and jump.find\_showers = "True". In
the subsequent MRS pipeline processing the background subtraction was performed in
a pixel-by-pixel fashion\footnote{see descriptive notes in section 4 of
https://github.com/STScI-MIRI/MRS-ExampleNB/blob/main/Flight\_Notebook1/MRS\_FlightNB1.ipynb}, and in the MRS Spec3Pipeline call outlier\_detection was set to "True", with an
outlier\_detection.kernal\_size = "11 1", and an outlier\_detection.threshold\_percent =
99.5. For both NIRSpec and the MRS the IFU Cube Build parameter,
cube\_build.weighting = "drizzle".
Asteroid spectra were extracted from each final spectral-spatial data cube which were "drilled"
along the cubes spectral axis using 1" effective circular aperture centered on the
photocenter of asteroid. In the case of the NIRSpec data, spectra were sigma clipped to
suppress residual "hot-pixels".

\subsubsection{NIRSpec Reflectance Spectrum} \label{subsec:nirspec reflectance spectrum}

\begin{figure}[!h]
    \centering
    \includegraphics[width=1\linewidth]{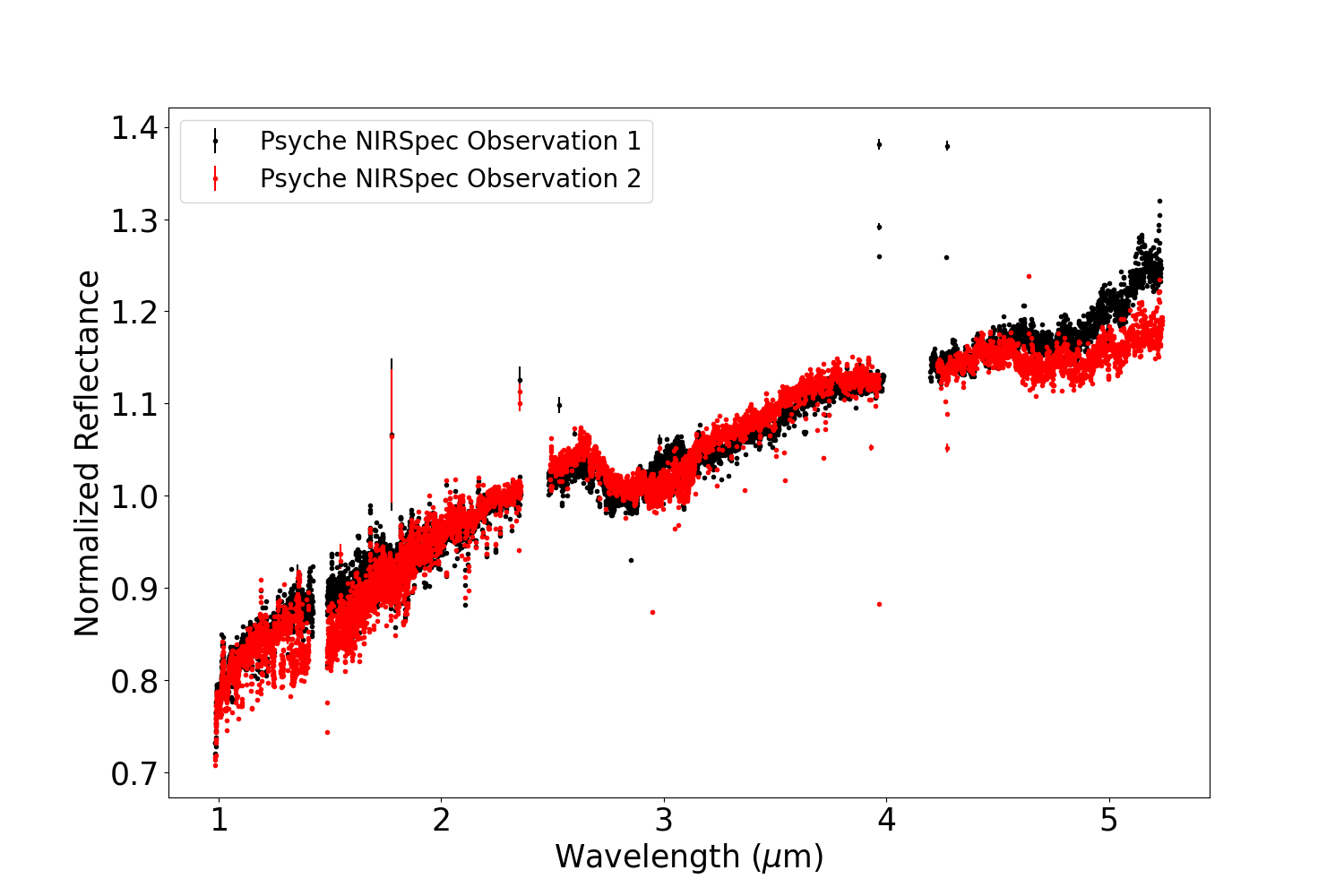}
    \caption{Psyche’s reflectance spectra from JWST NIRSpec observations. The data in black correspond to observation 1 and in red to observation 2. We omitted observation 1 G235H/F170LP mode data beyond 2.9 $\mu$m due to likely instrumental contamination.}
    \label{fig:Fig3Normrefl}
\end{figure}

We produced the reflectance spectrum by dividing the calibrated NIRSpec data by the solar spectrum obtained from the Planetary Spectrum Generator (PSG) using conditions matching the observations. The NIRSpec data also extend far enough into the infrared to be affected by thermal emission. We applied a tailored version of the Standard Thermal Model as established by \cite{1986Icar...68..239L} modified with some features from the Near Earth Asteroid Thermal Model (NEATM; \cite{1998Icar..131..291H}) to correct for the effect of thermal emission across the NIRSpec data (1.10 - 5.23 $\mu$m). The thermal model differs from the NEATM in that it does not account for night-side thermal emission, which may be neglected for phase angles as low as those of our observations (13.8 - 16.9\degree) and has been applied previously in \cite{2022PSJ.....3..153R} and \cite{2023PSJ.....4..214R}.  The model integrates various parameters including physical attributes such as radius, albedo, and emissivity, alongside observational parameters like the distances to the Sun and Earth, and the phase angle. The only free parameter of the model was the "beaming parameter" ($\eta$), a variable that encapsulates multiple factors influencing the asteroid's temperature. Determining the correct $\eta$ value is critical as it directly influences the extent of thermal flux removal, with each value suggesting a specific continuum behavior. We note, however, that the peak of Psyche’s thermal emission is well beyond 6 $\mu$m, and therefore thermal flux removal cannot create a false absorption feature contained within the bounds of the datasets. We assumed a linear continuum and iteratively adjusted $\eta$ values until finding one that aligned with the expected 3.75 $\mu$m reflectance ratio ($\eta$ = 0.8) using extrapolated data from shorter wavelengths without a thermal flux contribution. The thermally corrected reflectance spectrum is shown in Figure \ref{fig:Fig3Normrefl}. 

\FloatBarrier
\FloatBarrier

\subsubsection{MIRI Emission Spectrum} \label{subsec:nirspec reflectance spectrum}
We calculated Psyche's thermal emission for the MIRI dataset via a thermophysical model (TPM) code \citep{1996A&A...310.1011L}, \citep{1997A&A...325.1226L}, \citep{1998A&A...332.1123L}, \citep{1998A&A...338..340M}, and \citep{2002A&A...381..324M}. We used the latest available spin-shape solution \citep{2021PSJ.....2..125S} which is based on a wide range of radar measurements, plane-of-sky ALMA imaging, and adaptive optics images from Keck and the VLT. For the surface temperature calculations and the flux predictions, we used true solar insolation and JWST observing geometry for the two different MIRI epochs. Critical thermophysical properties like surface emissivity, thermal inertia or roughness are all based on a detailed TPM study \citep{2022A&A...659A..38R} which allowed us to interpret a wide range of thermal-IR measurements taken during the last 40 years. From that study, we took the best-fit thermal inertia of 50 s\textsuperscript{1/2} K\textsuperscript{-1}m\textsuperscript{-2 } and a low level of surface roughness for the interpretation of the MIRI measurements. We generated the solar spectrum for the MIRI observations in PSG as well and corrected for solar flux contributions by subtracting the solar flux from the total observed flux ($\sim
$0.02 Jy near 6 $\mu$m corresponding to $\sim$2\% of the total flux). We then divided this solar subtracted MIRI flux by the TPM to produce the MIRI emission spectrum. We then fit a line using data from 5.75 - 5.90 $\mu$m and 6.1 - 6.33 $\mu$m to detrend the data. To increase the signal-to-noise while retaining sufficient resolution to assess the presence of a 6-$\mu$m feature with an expected width of $\sim$100 nm we binned the data to 20 nm. 

\subsubsection{Near-Infrared Observations of Psyche} \label{subsec:nirspec observations}
\begin{figure}[!h]
    \centering
    \includegraphics[width=1\linewidth]{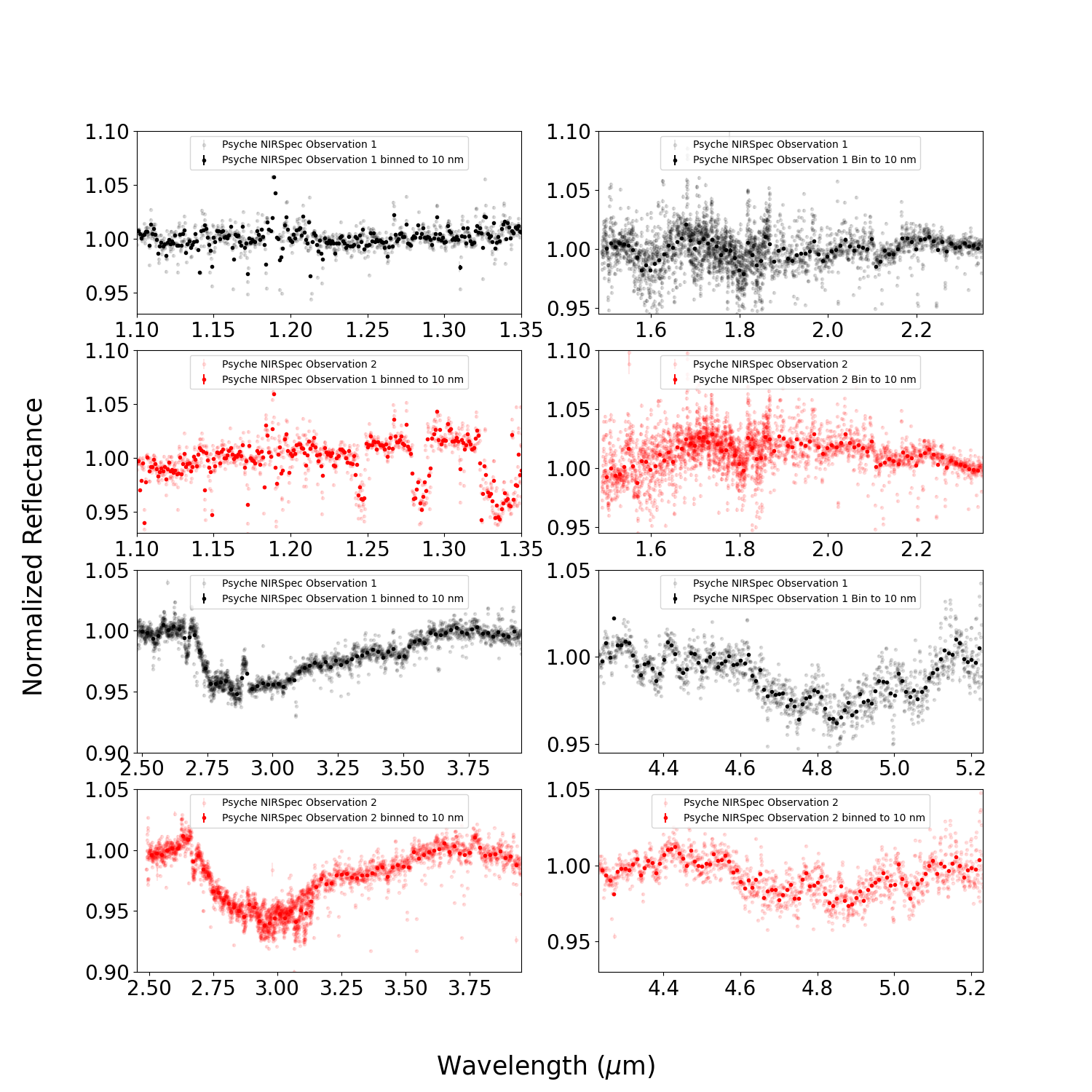}
    \caption{Psyche’s reflectance spectra from JWST NIRSpec observations. The data in black correspond to observation 1 and in red to observation 2. The darker points correspond to the binned data used to assess spectral feature properties. We omitted observation 1 G235H/F170LP mode data beyond 2.9 $\mu$m due to likely instrumental contamination.
}
    \label{fig:Fig4NormReflCont}
\end{figure}

After producing the NIRSpec reflectance spectrum, we split the NIRSpec observations into wavelength ranges of 1.10 - 1.35 $\mu$m, 1.48 - 2.35 $\mu$m, 2.48 - 3.95 $\mu$m, and 4.23 - 5.23 $\mu$m. These ranges represent the continuously available NIRSpec data (i.e., without gaps) over which we could fit a linear continuum to produce a normalized reflectance spectrum for each range. We note that the NIRSpec data begin at $\sim$1 $\mu$m, but we selected a cutoff of 1.1 $\mu$m as we believe the 1.0 - 1.1 $\mu$m range is dominated by instrumental effects. We divided each of these wavelength ranges by separate fit linear continuums to produce normalized reflectance spectra to identify absorption features. We first fit a continuum using only portions of the beginning and end of the wavelength range, then once we identified potential spectral features we fit a new continuum over the entire wavelength range with those spectra features masked. We binned the data between 1 and 20 nm depending on the data resolution prior to fitting a gaussian to any identified features. We also omitted G235H/F170LP mode data beyond 2.9 $\mu$m for observation 1 as these data appeared to be dominated by a potential instrumental effect with flux values largely inconsistent with the G395H data from the same observation and G235H data from observation 2. The resulting normalized reflectance spectra and corresponding spectra binned to 10 nm for each wavelength range are shown in Figure \ref{fig:Fig4NormReflCont}. 

\FloatBarrier
\FloatBarrier

We measured the band depth of the 3-$\mu$m feature in two ways. First, we applied a gaussian fit to the 3-$\mu$m region. Second, we took the centroid of the flux between 2.7 and 3.1 $\mu$m and estimated the band strength by taking the mean of the observed flux at the estimated band center subtracted from the the mean continuum flux at the estimated band center. Third, we set the band center as the minimum value of the data binned to a wavelength of 10 nm and measured the band depth as the observed minimum flux subtracted from the mean continuum flux at this band center. For observation 1 this was repeated with the spike between 2.87 and 2.91 $\mu$m excluded from the analysis. The results of these various methods are in Table \href{}{\ref{tab:Table1}}

\subsubsection{Mid-Infrared Observations of Psyche} \label{subsec:MIRI observations}

We subtracted the solar contribution from the MIRI data and then binned the data to a wavelength resolution of 20 nm. We then divided the data by a spin-shape thermophysical model and fit a line that masked the 6-$\mu$m feature region (5.9 - 6.1 $\mu$m) and longward of 6.33 $\mu$m to remove the continuum and produce a normalized emission spectrum. We then fit a Gaussian to the data in the 6-$\mu$m feature region and calculated the standard deviation around the feature region to report an upper limit on the detectable amount of water potentially present.

\section{Results}

The principal objective of this study is to investigate the presence of 3 and 6-$\mu$m features on Psyche's surface to evaluate the corresponding abundance and heterogeneity of hydration across Psyche's surface. Alongside this main goal, we have also mapped additional spectral features detailed in Table \ref{tab:Table3_AdditionalFeatures}. In the following sections, we describe our approach for identifying and characterizing spectral features from the NIRSpec and MIRI observations.  

\subsection{3-$\mu$m Feature}
\begin{figure}[!h]
    \centering
    \includegraphics[width=1\linewidth]{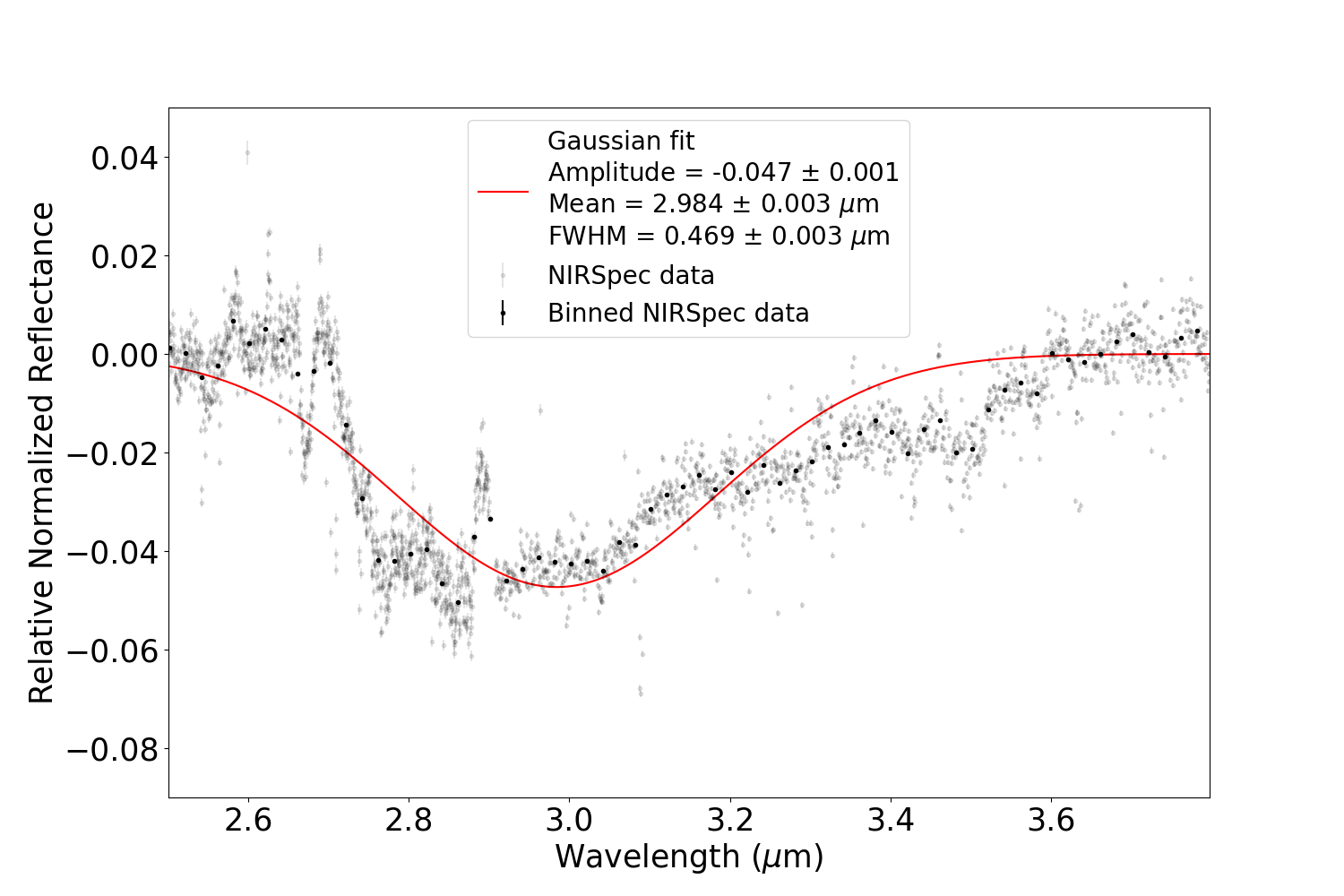}

    \includegraphics[width=1\linewidth]{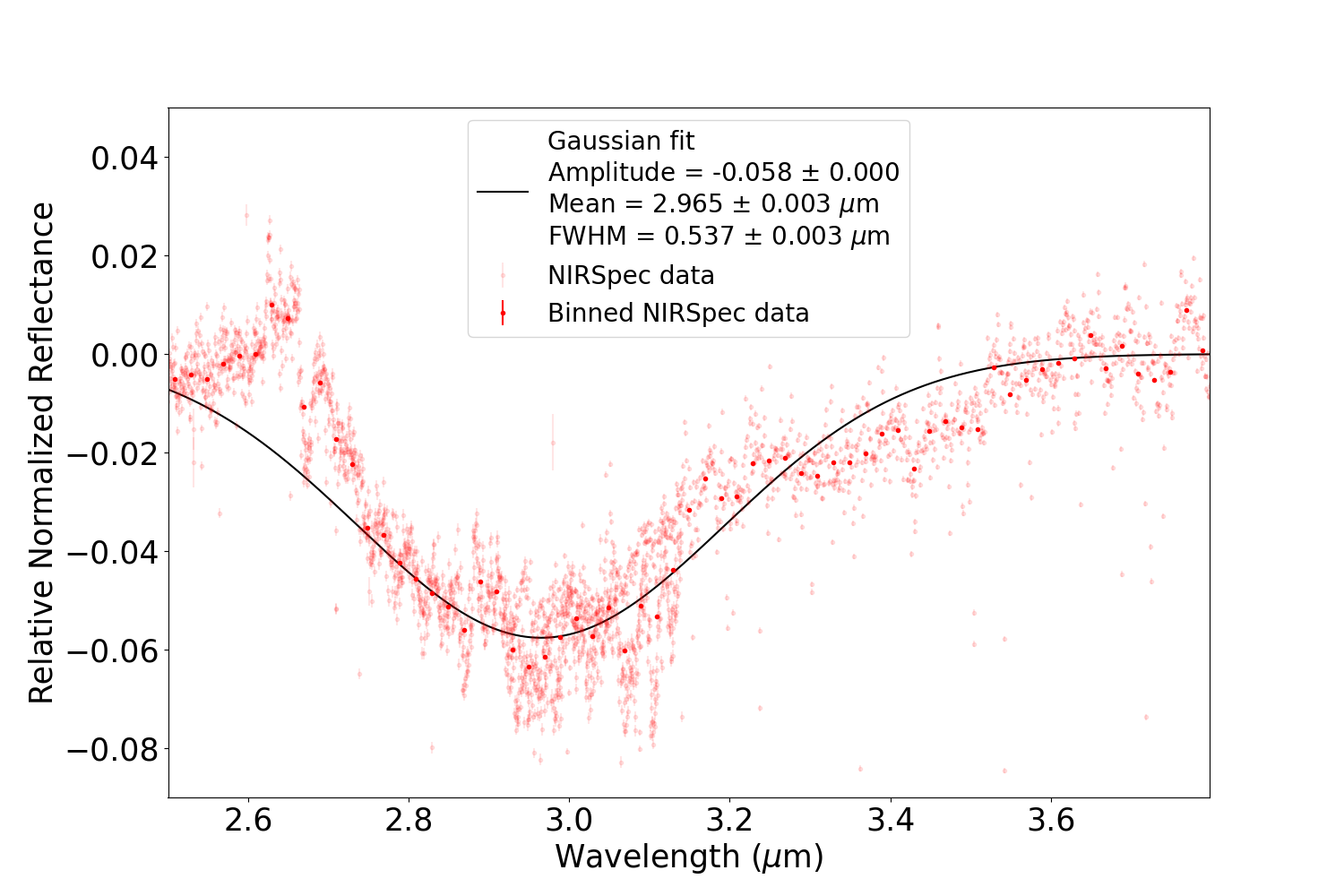}
    \caption{Relative normalized reflectance spectrum of Psyche near 3-$\mu$m (Top) observation 1 (Bottom) observation 2. }
    \label{fig:Fig5_3micgaussfit}
\end{figure}

We observed a 3-$\mu$m feature associated with OH, and potentially H\textsubscript{2}O, in each observation and applied a Gaussian fit to the region to determine the feature band depth, band center, and width (Figure \ref{fig:Fig5_3micgaussfit}). The relative normalized reflectance spectra in Figure \ref{fig:Fig5_3micgaussfit} is the normalized reflectance subtracted from the mean of the continuum between 3.6 and 3.7 $\mu$m such that the average of the continuum would be 0. We compare these fit parameters to an estimation of the band center by taking the centroid of flux within the wavelength range of 2.7 to 3.1 $\mu$m and estimating the band depth taking the mean observed flux at the estimated band center and subtracting this from the mean continuum flux at the estimated band center. 

\FloatBarrier
\FloatBarrier

The estimated band center for observation 1 using the centroid method was 2.89 $\mu$m with a band depth estimate of 4.3\%. The estimated band center from observation 2 from the centroid method was 2.90 $\mu$m with a band depth estimate of 4.9\%. These same parameters from the Gaussian fit for observation 1 were a band center and depth respectively of 2.92 $\mu$m and 4.5\% and for observation 2 were 2.96 $\mu$m and 5.8\%. For observation 1 we also applied the Gaussian fit method to the data while masking the spike between 2.87 and 2.91 $\mu$m resulting in a band center of 2.98 and a band depth of 4.9\%. Taking the minimum of the data binned to 10 nm for observation 1 resulted in a band center of 2.86 $\mu$m with a band depth of 5.1\% and for observation 2 a band center of 2.95 with a band depth of 6.0\%. The band depths calculated from the variety of approaches are given in Table \ref{tab:Table1}. The reported errors for the band depth measurements from the centroid method are assigned from the standard deviation of the data from 3.56 - 3.90 $\mu$m. When varying the linear fit by the standard error of the fit slope we found that the variation of the estimated band depth was within the standard deviation of this baseline and when applying the band depth error calculation method of \cite{2017AJ....153...31T} the error was on the order of $\sim$0.001\% and therefore insignificant compared to variations that may arise from the choice of a linear fit to the continuum and standard deviation of the data.

\begin{table}[!h]
\centering
\caption{ 3-$\mu$m feature property estimates for each NIRSpec observation using centroid and Gaussian fit methods applied to the data for each observation and for observation 1 the sharp feature between 2.87 and 2.91 $\mu$m removed.}
\label{tab:Table1}
\begin{tabular}{| l | l | l | l |}
\hline
\textbf{NIRSpec Observation} & \textbf{Method} & \textbf{Band Center ($\mu$m)} & \textbf{Band Depth (\%)} \\
\hline
1 & Centroid & 2.89 & 4.3 ± 0.3 \\
 & Centroid with central spike removed & 2.89 & 4.4 ± 0.3 \\
 & Gaussian & 2.92 & 4.5 ± 0.2 \\
 & Gaussian with central spike removed & 2.98 & 4.9 ± 0.2 \\
 & Minimum of binned & 2.86 & 5.1 ± 0.3 \\
\hline
2& Centroid & 2.90 & 4.9 ± 0.4 \\
 & Gaussian & 2.96 & 5.8 ± 0.2 \\
 & Minimum of binned & 2.95 & 6.0 ± 0.4 \\
\hline

\end{tabular}

\end{table}

\begin{figure}[!h]
    \centering
    \includegraphics[width=.75\linewidth]{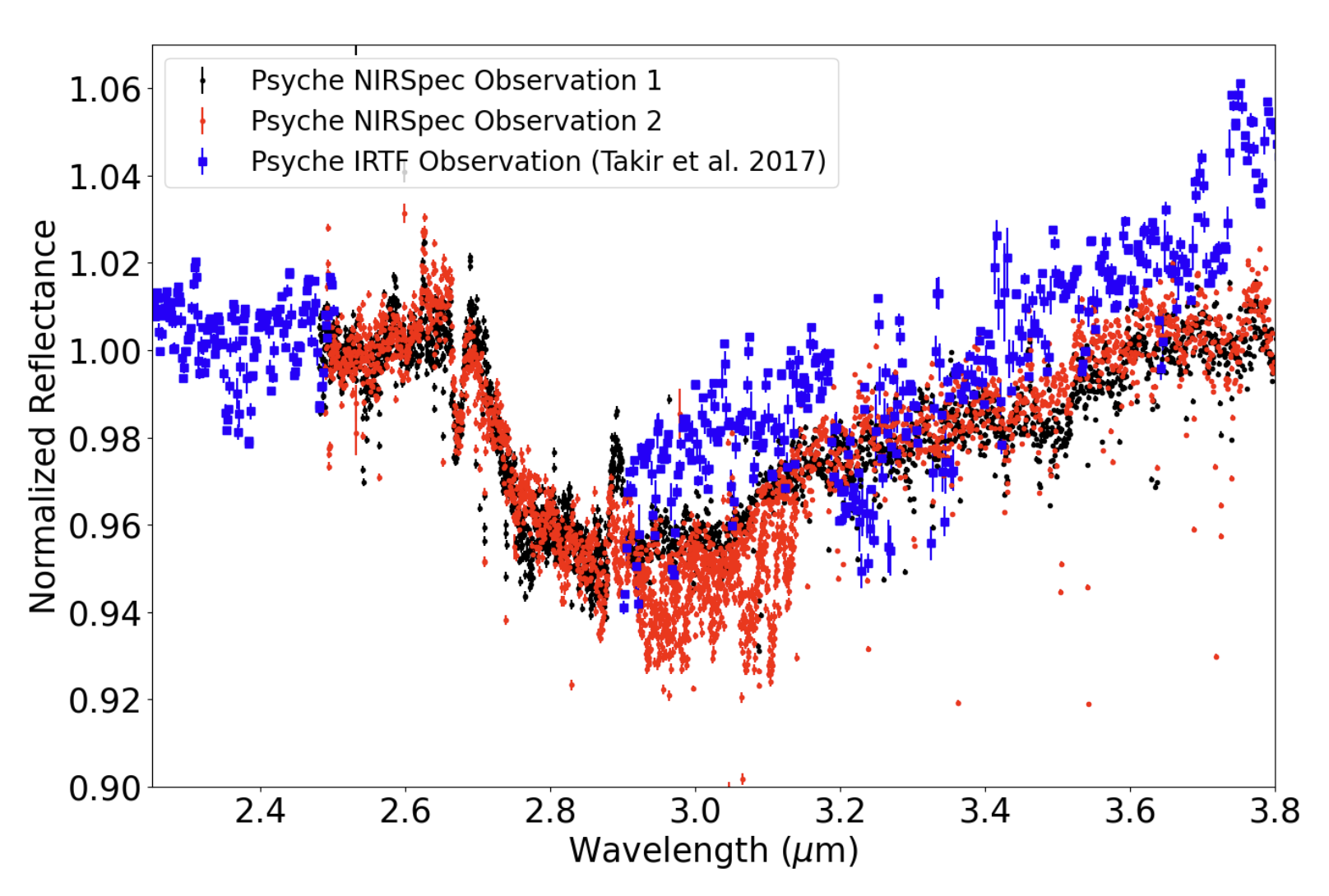}
    \caption{Relative normalized reflectance spectrum of Psyche near 3-$\mu$m for NIRSpec observation 1 (black) observation 2 (red) and IRTF from \cite{2017AJ....153...31T} (blue).}
    \label{fig:Fig6IRTF3mic}
\end{figure}

The band depths estimated from the observations by \cite{2017AJ....153...31T} range between 2.72 - 3.26\%, the center of these bands was not measurable due to the lack of available data resulting from atmospheric opacity at the band center and therefore the authors assumed a center of 3 $\mu$m (see Figure \ref{fig:Fig6IRTF3mic})

\FloatBarrier
\FloatBarrier

\begin{figure}[!h]
    \centering
    \includegraphics[width=.75\linewidth]{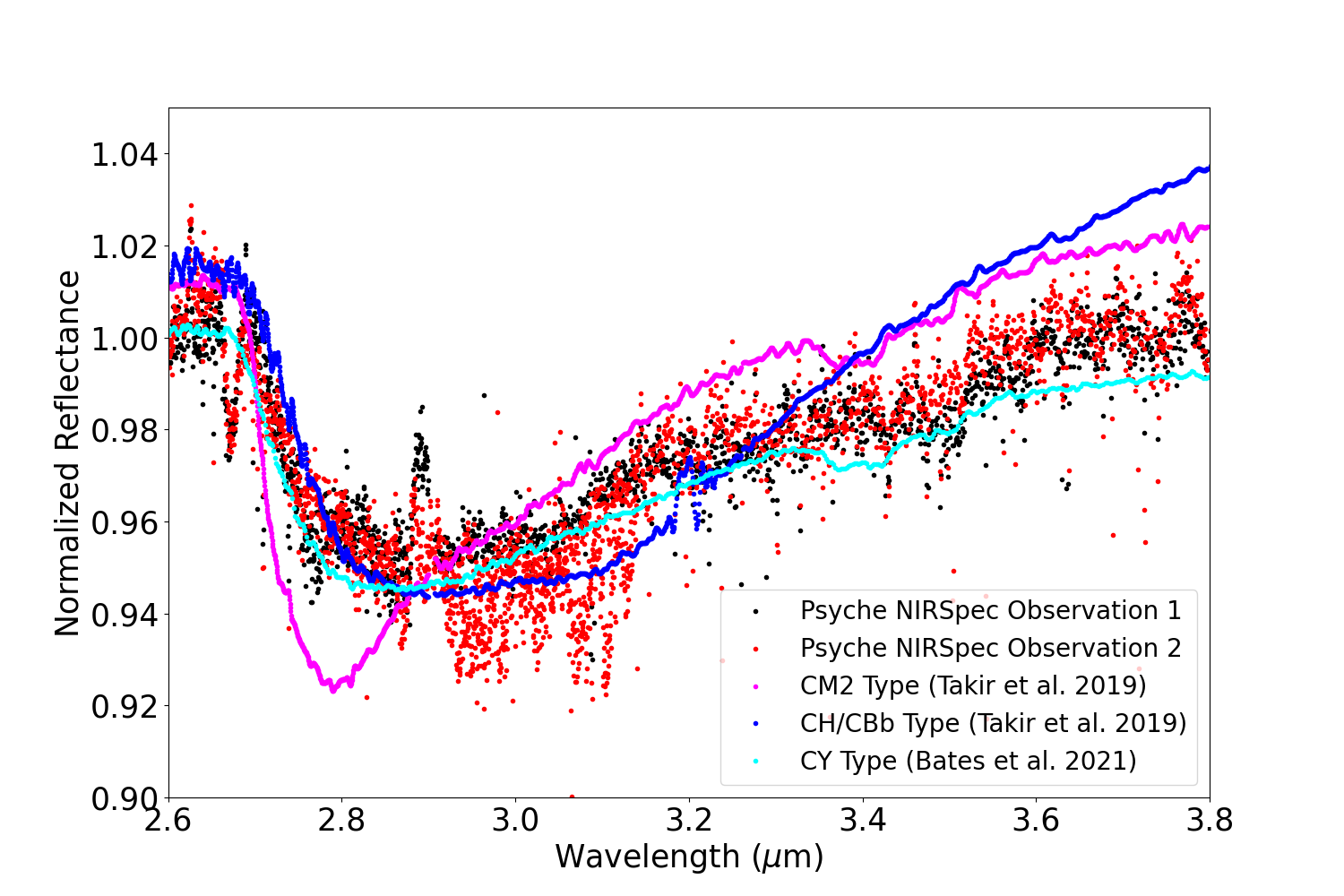}
    \caption{Normalized reflectance spectrum of Psyche near 3-$\mu$m from observation 1 (black), observation 2 (red), normalized reflectance spectra for laboratory data from CM2- (magenta), CH/CBb- (blue), and CY-type (cyan) carbonaceous chondrites. The 3-$\mu$m feature on Psyche most closely resembles CBb-type chondrites. }
    \label{fig:Fig7}
\end{figure}

We estimate that the band depth of the 3-$\mu$m feature is between 4.3 and 4.7\% for observation 1 and 5.3 and 6.0\% for observation 2. The differences in measured feature strength from JWST and IRTF observations further suggest heterogeneity of the hydration abundance across Psyche’s surface by a few percent. We also compared the normalized reflectance spectrum of Psyche near 3 $\mu$m to laboratory spectroscopic measurements of carbonaceous chondrites from \cite{2019Icar..333..243T} and \cite{2021JGRE..12606827B} (Figure \ref{fig:Fig7}).

\FloatBarrier
\FloatBarrier

\begin{figure}[!h]
    \centering
    \includegraphics[width=.75\linewidth]{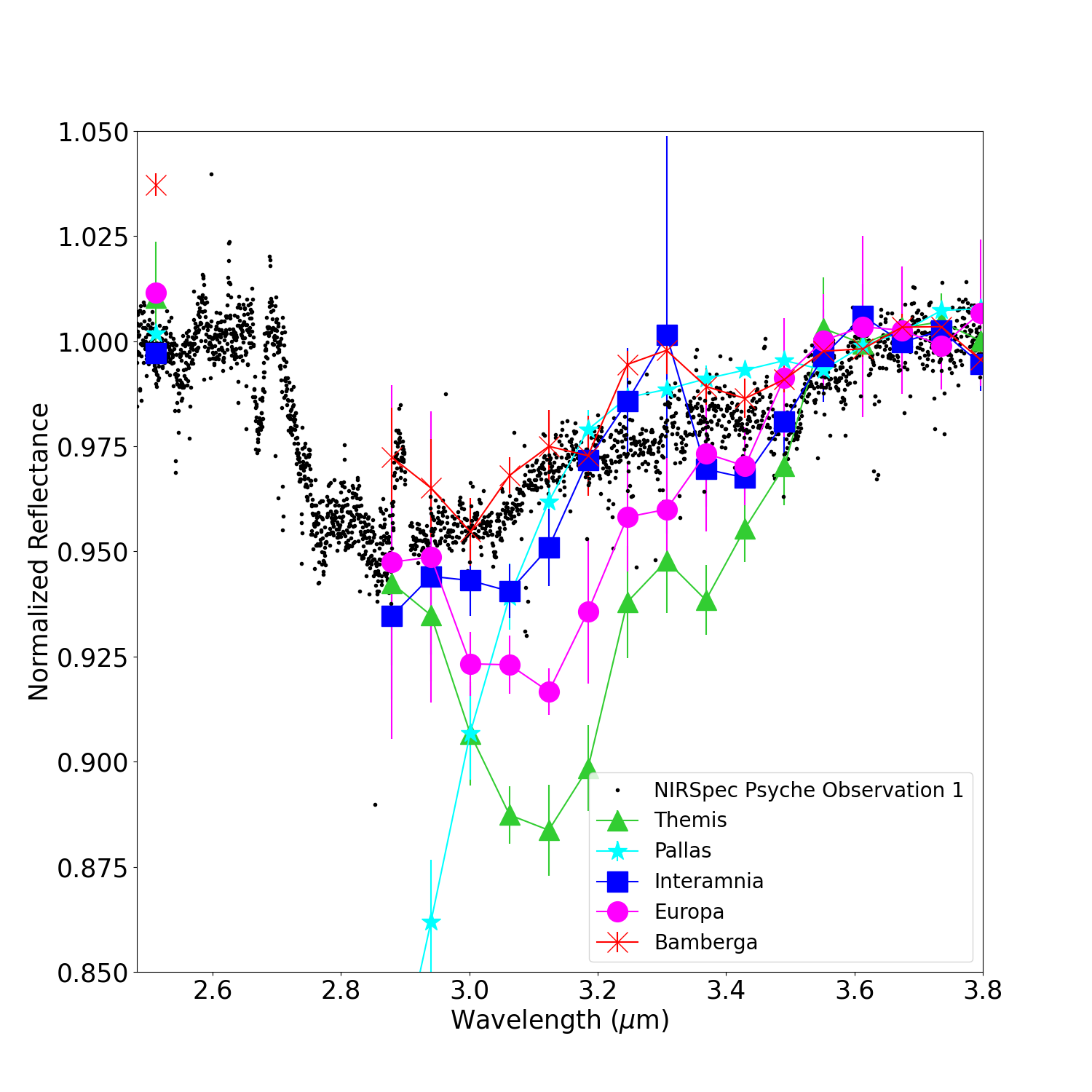}
    \caption{Normalized reflectance spectrum of Psyche near 3-$\mu$m from observation 1 (black), normalized reflectance spectra for asteroid data from near-infrared observations of Themis (green), Pallas (cyan), Intermnia (blue), Europa (magenta), and Bamberga (red) with data points corresponding to a binned average using 25 bins and errors corresponding to the standard deviation across the bin. }
    \label{fig:Fig8}
\end{figure}

Figure \ref{fig:Fig7} shows the normalized reflectance spectrum of Psyche near 3-$\mu$m compared with laboratory spectra of CM2-, CH/CBb- and CY-type carbonaceous chondrites. The shape of the 3-$\mu$m feature more closely resembles CY and CH/CBb type chondrites. We include the CM2 data as a common example of sharp 2.7-$\mu$m features typically associated with OH present both in the chondrites and asteroids e.g., Bennu from \cite{2023Icar..40015563C} that do not match the feature shape observed on Psyche. The CH/CBb type chondrite Isheyevo shown in blue in Figure \ref{fig:Fig7} was also found to be a good match for the vis-NIR spectra of Psyche from \cite{2023E&SS...1002694D}. We also found that partial 3-$\mu$m features observed on asteroids (324) Bamberga and (704) Interamnia, both considered to have either sharp type or not sharp type 3-$\mu$m features depending on the observation, most closely matched the shape of  Psyche’s 3-$\mu$m feature when performing a sum of square difference comparisons between Psyche and the asteroid spectra provided in \cite{2022PSJ.....3..153R} (Figure \ref{fig:Fig8}). The sum of square differences between Psyche’s spectra and that of (324) Bamberga and Interamnia were as low as 0.11. For comparison, the lowest values for (24) Themis, (52) Europa, and (2) Pallas were 2.18, 0.81, and 1.21 respectively. 

\FloatBarrier
\FloatBarrier

It is possible that Psyche’s 3-$\mu$m band shape may have been a 2.7-$\mu$m sharp type that shifted closer to 2.9-$\mu$m taking on a more rounded shape as a result of thermal metamorphism (as described in \citep{2021JGRE..12606827B}) . Psyche’s 3-$\mu$m band is also a close match to the stage III CY 980115 chondrite from \citep{2021JGRE..12606827B} that contains $\sim$8 vol\% secondary olivine, $\sim$2 vol\% magnetite, $\sim$11 vol\% Fe-sulfide, and $\sim$79 vol\% dehydrated phyllosilicate. Stage III chondrites are characterized by low degrees of aqueous alteration and unequilibrated mineral assemblages with anticipated peak temperatures of the parent body around 600\degree C. From \cite{2021JGRE..12606827B}, 3-$\mu$m features for chondrites in stage III samples show features between 2.88 and 3.02 $\mu$m (as we see on Psyche),though the position of the feature is not influenced by degree of aqueous alteration and therefore offers no information on phyllosilicate abundance.

\subsection{6-$\mu$m Feature }
\begin{figure}[!h]
    \centering
    \includegraphics[width=.75\linewidth]{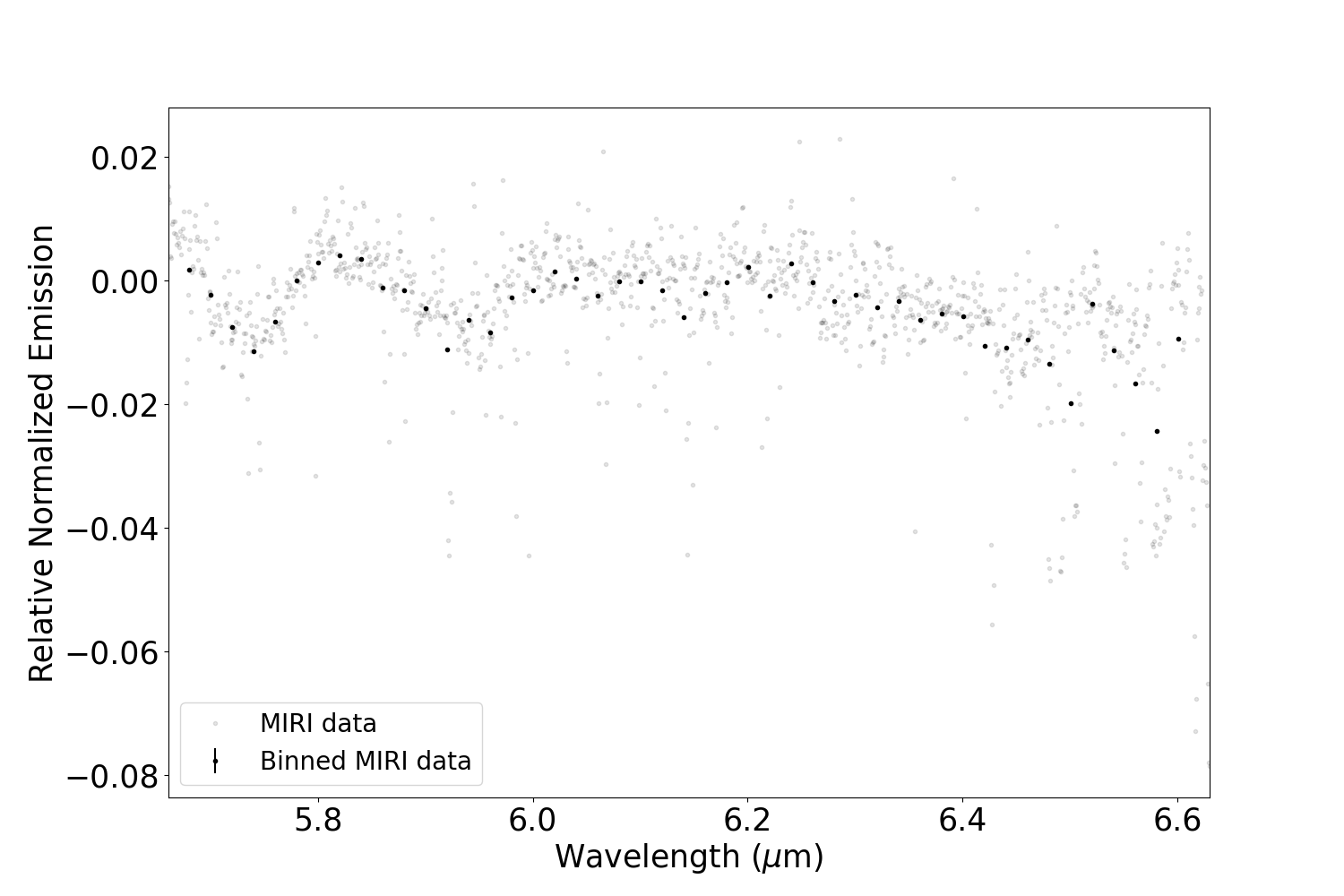}

        \includegraphics[width=.75\linewidth]{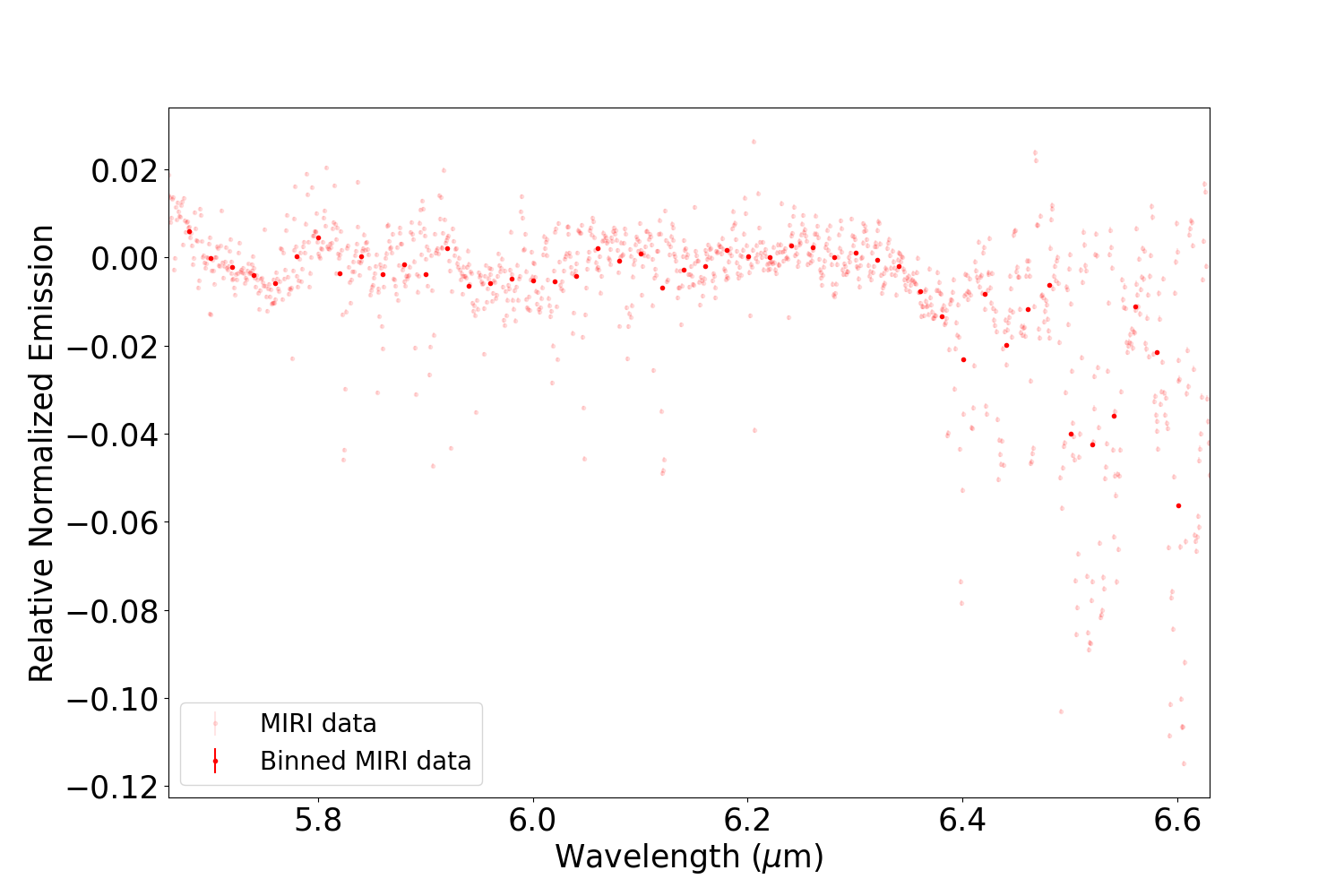}
        \caption{Relative normalized emission spectrum of asteroid (16) Psyche produced by dividing the MIRI channel 1 flux by the TPM then dividing by a linear fit for (Top) observation 1 (Bottom) observation 2. The data are binned to a resolution of 20 nm. \textbf{We do not detect any 6-$\mu$m emission feature in either observation}.}
        \label{fig:6micronfits_Figure9}
   
\end{figure}

We did not detect any definitive 6-$\mu$m feature that would have potentially been associated with H\textsubscript{2}O in either MIRI observation. Figure \ref{fig:6micronfits_Figure9} shows the relative normalized emission for each observation. The normalized emission was subtracted from the mean of the continuum between 6.15 and 6.22 $\mu$m such that the average of the continuum would be 0. 

\FloatBarrier
\FloatBarrier

The data reduced using a previous version of the pipeline (v.1.11.3 with pmap 0994) showed a peak exceeding ~1\% above the baseline in observation 1 and this was no longer observed after reducing the data with a more recent version of the pipeline (v.1.14.0 with pmap 1241). The standard deviation of data near where the band may be present (from 6.15 - 6.22 $\mu$m) is 0.0037 or 0.37\%. We also don't anticipate that the large scatter of up to 0.1 or 10\% between 6.4 and 6.6 $\mu$m is due to any spectral features and therefore suggests an instrumental artifact. We therefore do not report a definitive detection of water on Psyche, but cannot completely rule out its absence, particularly if it is as low as a 0.4 - 1\% level which may be anticipated based on values observed on the Moon and at Itokawa of between 1 - 5\%. To provide a reasonable upper limit on the amount of water that may be present at Psyche but would potentially not be observable as a result of pipeline limitations and instrumental effects, we report this upper limit conservatively at 0.4\%. The water abundance from the strength of the 6-$\mu$m feature may be calculated using the equation from \cite{2021NatAs...5..121H} \[A = 9394 \cdot D\textsuperscript{2} + 9594\cdot D\] 
where A is abundance in ppm and D is the band strength. Using a band strength of 0.4\% this amounts to an upper limit for the water abundance of 39 ppm. 

The possible interpretations for a non-detection include: 
1. Water is present but at an amount that the instrument was not sensitive to, which would likely be below 0.4\% or 39 ppm. 
2. Water is present above this level, up to potentially 10\%, but issues with the calibration pipeline have suppressed the feature. 
3. Water is not present at all in the hemisphere observed by these JWST observations. 
4. Water is present but other spectral absorption features near 6-$\mu$m may be present as well. 

Of these possibilities, interpretation 1 seems most likely given that water abundance estimates of 100 - 500 ppm have been measured at the Moon \citep{2021NatAs...5..121H} and on near-earth asteroids such as Itokawa \citep{2021NatAs...5.1275D} (corresponding to 1 - 5\% band strengths), but Psyche's farther distance from the Sun leads to a correspondingly lower exposure to solar wind implantation. The hydrogen ions from the solar wind are thought to interact with the surface of these airless bodies to produce water, and Psyche's reduced exposure to this solar wind may lead to a correspondingly reduced amount of water produced. As a result we would not necessarily anticipate significantly higher water abundance than on the surface of the Moon, though we cannot definitively rule this out due to instrumental effects that are apparent in the data and due to deviations in spectral features that arise as a result of updates to the calibration pipeline. Additionally, the relationship between hydration and the solar wind may be further complicated by cooler temperatures for bodies farther away from the Sun potentially increasing the retention of water even though the body has less access to solar wind hydrogen ions.  We also note that olivine has an emission feature near 6-$\mu$m \citep{2020GeoRL..4789151K} and other minerals may have absorption features in emissivity near this wavelength as well. This points to possible degeneracies in interpretations of the strength of the 6-$\mu$m band. The lack of any water at all would indicate that the hydration producing the 3-$\mu$m band is entirely due to OH, which is possible, though a limited amount of water under the detection threshold of the instrument is more likely given the exposure of the asteroid to hydrogen ions that would likely interact with the surface in a similar way to other airless bodies. 

We also note that \cite{2018Icar..304...74R} identify a relationship in non-carbonaceous asteroids between hydrogen abundance and the band depth at 2.95 $\mu$m where a 2\% band depth corresponds to $\sim$100 pm of hydrogen. We are reluctant to make any estimates of specific amounts of hydrogen abundance inferred from this relationship based on the largest band depths observed at Psyche because the relationship is likely nonlinear and is unknown. However, observations of other airless bodies' 3-$\mu$m bands with depths similar to what is observed at Psyche, e.g. at Eros and Ganymed, have had hydrogen estimates placed at 250 - 400 ppm \citep{2018Icar..304...74R}. Other sources have correlated a 2\% absorption feature on the Moon to $\sim$1000 ppm water \citep{2009Sci...326..562C}. We do not claim that a similar amount of water exists on Psyche, particularly because no 6-$\mu$m emission feature uniquely associated with water was observed, but point out that the relationship between hydrogen, OH and H\textsubscript{2}O abundance from 3-$\mu$m feature properties alone may be insufficient and limitations in degeneracies of interpretation of the 6-$\mu$m feature generally may further complicate inferred water abundance. 

\subsection{Additional Spectral Features}

We identified several other spectral features by visual inspection and further assessed these by fitting their parameter properties. The identified absorption features in NIRSpec observations are located at 1.25 and 1.28 $\mu$m in observation 2, 1.59 $\mu$m in observation 1, and 2.66 and 4.8 $\mu$m in both observations. The identified absorption features in the MIRI emission spectrum are located at 5.74 and 5.93 $\mu$m in observation 1 and 5.75 and 5.98 $\mu$m in observation 2. We fit Gaussians to these features to determine their properties (see Appendix A. Supplemental Figures) and recorded these in Table \ref{tab:Table3_AdditionalFeatures}. The data with the 1.25, 1.28, and 1.34 $\mu$m features were binned to 1 nm, the data with the 1.59 $\mu$m feature were binned to 10 nm, the data with the 2.66 $\mu$m feature were binned to 1 nm, and the data with the 4.8 $\mu$m feature and the MIRI data were binned to 20 nm. 

\FloatBarrier
\begin{table}[!h]
\centering
\caption{Properties of additional infrared absorption features detected on Psyche. }
\label{tab:Table3_AdditionalFeatures}
\begin{tabular}{| l | l | l | l | l |}
\hline
\textbf{Band Center ($\mu$m)} & \textbf{Instrument} & \textbf{Observation} & \textbf{Band Depth (\%)} & \textbf{FWHM ($\mu$m)} \\
\hline
1.25 ± 0.00 & NIRSpec & 2 & 4.3 ± 0.9 & 0.004 ± 0.00 \\
\hline
1.28 ± 0.00 & NIRSpec & 2 & 5.6 ± 0.5 & 0.009 ± 0.000 \\
\hline
1.34 ± 0.00 & NIRSpec &  2 &  6.3 ± 0.2 & 0.023 ± 0.001 \\
\hline
1.59 ± 0.00 & NIRSpec &  1 &  2.3 ± 0.2 & 0.059 ± 0.002 \\
\hline
2.67 ± 0.00 & NIRSpec &  1 &  2.6 ± 0.2 & 0.013 ± 0.000 \\
\hline
2.67 ± 0.00 & NIRSpec & 2 &  2.9 ± 0.1 & 0.008 ± 0.001 \\
\hline
4.84 ± 0.01 & NIRSpec &  1 &  2.8 ± 0.2 & 0.331 ± 0.014 \\
\hline
4.81 ± 0.02 & NIRSpec &  2 &  2.0 ± 0.2 & 0.443 ± 0.025 \\
\hline
5.74 ± 0.00 & MIRI &  1 & 1.1 ± 0.4 & 0.048 ± 0.003 \\
\hline
5.75 ± 0.00 & MIRI &  2 & 0.7 ± 0.4 & 0.034 ± 0.003 \\
\hline
5.93 ± 0.00 & MIRI &  1 & 0.9 ± 0.4 & 0.067 ± 0.004 \\
\hline
5.98 ± 0.00 & MIRI &  2 & 0.6 ± 0.4 & 0.105 ± 0.009 \\
\hline

\end{tabular}
\end{table}
\FloatBarrier

The 1.25, 1.28, 1.34, 1.59, and 2.67 $\mu$m absorption features observed are quite narrow (FWHMs 0.004 - 0.023 $\mu$m), with strengths that are unlikely to have gone undetected by previous observations in these wavelengths (we were unable to find these features in previous near-IR observations of Psyche described in \cite{2008Icar..195..206O}). We also report absorption features in emissivity near 5.75 and 5.95 $\mu$m. We are uncertain whether any of these correspond to real features but we include them for comparative purposes as more JWST asteroid and laboratory observations become available. We compared the emissivity spectra to that of olivine and orthopyroxene and found no similarities. However, the broad 4.81 $\mu$m absorption feature in both observations may be associated with an Si-O bending feature present with silicate material providing additional evidence for the presence of silicate material across Psyche’s surface. 

\section{Discussion}

JWST observations of asteroid (16) Psyche offer critical insights into the asteroid's surface composition and its geological history while providing context for upcoming \textit{Psyche} mission in-situ observations. The presence of 3-$\mu$m features and the lack of a 6-$\mu$m feature in Psyche's spectrum has profound implications for understanding the asteroid's origin, evolution, and the broader narrative of water distribution in the early Solar System.

The detection of the 3-$\mu$m absorption feature associated with hydroxyl or water molecules in our NIRSpec observations suggests that Psyche's surface is not exclusively metallic but hosts widespread hydrated material. Laboratory studies of the 3-$\mu$m features in meteorites show trends in band shape and center with the degree of aqueous alteration \citep{2013M&PS...48.1618T}, \citep{2014Icar..229..263B}, \citep{2019M&PS...54.2652G}, and \citep{2019Icar..333..243T}. The nature of the 3-$\mu$m feature observed on Psyche is not comparable to the sharp shape centered at $\sim$2.7-2.8 $\mu$m seen in CI-, CM-, and CR- type carbonaceous chondrites (i.e., OH in phyllosilicates). Instead, the band shape and center appear consistent with those in CY-, CH- and CB-type carbonaceous chondrites. The CHs are rich in metal (20\%; \cite{2003GeCAS..67R.237K}, \cite{2008M&PS...43..915I}) and the 3-$\mu$m band could be attributed to the presence of iron oxyhydroxides (e.g., FeO (OH), rust: FeO (OH), H\textsubscript{2}O). Previous studies have proposed that CBs formed in a vapor-melt plume produced by a giant impact, and the presence of CB-like material on Psyche’s surface could be the result of a giant impact removing Psyche’s mantle \citep{2017AJ....153...29S}. The variability in the strength of the hydration features across our observations implies a heterogeneous distribution of hydrated minerals, which could result from impacts, suggesting a complex surface history involving hydrated impactors with CHs and CBs as reasonable impactor candidates. 

The lack of a definitive detection of a 6-$\mu$m emission feature on Psyche indicates that Psyche's hydration is dominated by OH. Based on the standard deviation of the data around this feature location we set an upper limit of the potential presence of water that is below our detection limit at an abundance of 39 ppm. This value is less than half of and up to an order of magnitude lower than the molecular abundance of water on the Moon ($\sim$100 - 400 ppm \cite{2021NatAs...5..121H}), and on S-type asteroids that were previously considered anhydrous ($\sim$450 ppm \citep{2024PSJ.....5...37A}). However, the presence of water at an amount up to an order of magnitude lower than at the Moon is reasonable considering the solar wind flux at the Moon is 7 cm\textsuperscript{-3} \citep{2006P&SS...54..132F} and at Psyche is 0.6 cm\textsuperscript{-3} \citep{2022ApJ...927..202C}, and we may expect that if water is produced mostly as the result of solar wind implantation the correspondingly lower flux of solar wind would produce less water. Additionally, such a low upper limit of abundance lends further evidence that hydration is due to an exogenic source likely with some contribution from solar wind implantation but not yielding a significant amount of hydration due to water. 
 
 The range of possible interpretations highlights the enigmatic nature of Psyche's composition and inferred evolution. If Psyche’s hydration is endogenous, this supports the proposal that the asteroid may have formed beyond the snow line of the early solar system and later became implanted in the outer main belt \citep{2022SSRv..218...17E}. The presence of H\textsubscript{2}O and/or OH on Psyche could be attributed to either the delivery of hydrated material through impacts, space weathering processes, or the scenario that Psyche’s overall composition is not consistent with M-class asteroids (i.e., iron meteorites). Psyche could instead have the same composition as a P-class asteroid (with a somewhat higher albedo than typical P-class asteroids) or an E-class asteroid (with a lower albedo than a typical E-class asteroid). Additionally, the depth of the 3-$\mu$m absorption feature observed on Psyche is similar to that of other airless bodies (e.g., Eros and Ganymed; \cite{2018Icar..304...74R}) with hydrogen abundance estimates of 250 - 400 ppm. This estimated amount should be detectable by the \textit{Psyche }mission's Gamma Ray and Neutron Detection instrument and has implications for interpretations of the results from that instrument \citep{2024AGUA....501077D}. 

The\textit{ Psyche} mission will explore the asteroid's composition, internal structure, and geology to better understand the building blocks of planet formation. Our JWST observations provide a crucial pre-mission snapshot of Psyche's surface composition using a wavelength range outside the spacecraft’s capabilities. Understanding the distribution and abundance of hydrated minerals on Psyche will help to interpret the mission's data within the context of the asteroid's formation and evolutionary history. Our findings suggest that there are hydrated materials on Psyche’s north pole, potentially observable as silicates in chondrites that in addition to OH contain pyroxene which the \textit{Psyche }mission will be able to refute or verify through visual to near-infrared observations with the Multispectral Imager \citep{2023E&SS...1002694D}. 

The detection of hydration features on Psyche contributes to the growing body of evidence that M-class asteroids are a diverse population and that hydration is more widespread in the asteroid belt than previously thought. This has significant implications for our understanding of water delivery to the inner Solar System, potentially offering insights into the contribution of differentiated planetesimals (which are traditionally thought to be volatile-poor as a result of igneous differentiation) to the Earth's water budget. Psyche's complex surface composition, indicative of a rich geological history involving both metallic and hydrated components, challenges traditional asteroid classification schemes and calls for a reevaluation of our understanding of asteroid formation and evolution.

\section{Data Availability}
JWST data are publicly available from the Space Telescope Science Institute’s Mikulski Archive for Space Telescopes https://mast.stsci.edu/. The observations produced by the JWST 1731 observation program can be accessed via DOI at 10.17909/xmzs-f849. 

Reduced data used in this analysis and the data required to reduce them (i.e.,the PSG solar spectrum and thermal model outputs) are publicly available at Zenodo 10.5281/zenodo.12536821.

\section{Code Availability}
The Planetary Spectrum Generator used to generate the solar spectrum is at https:/psg.gsfc.nasa.gov/. The JWST science data calibration pipeline is at https://github.com/spacetelescope/jwst. The pipeline version used to calibrate the NIRSpec data was v1.11.3 \cite{bushouse_2023_8157276} and the pipeline version used to calibrate the MIRI data was v1.14.0 \cite{bushouse_2024_10870758}.

\section{Acknowledgements}
We thank the reviewers for their helpful feedback that led to substantial improvement of this manuscript. This work is based (in part) on observations made  with the NASA/ESA/CSA James Webb Space Telescope. The data were obtained from the Mikulski Archive for Space Telescopes at the Space Telescope Science Institute, which is operated by the Association of Universities for Research in Astronomy, Inc., under NASA contract NAS5-03127 for JWST. These observations are associated with program \#1731.



\bibliography{sample631}{}
\bibliographystyle{aasjournal}

\appendix

\section{Supplementary Figures}

\FloatBarrier
\begin{figure}[!h]
    \centering
    \includegraphics[width=.5\linewidth]{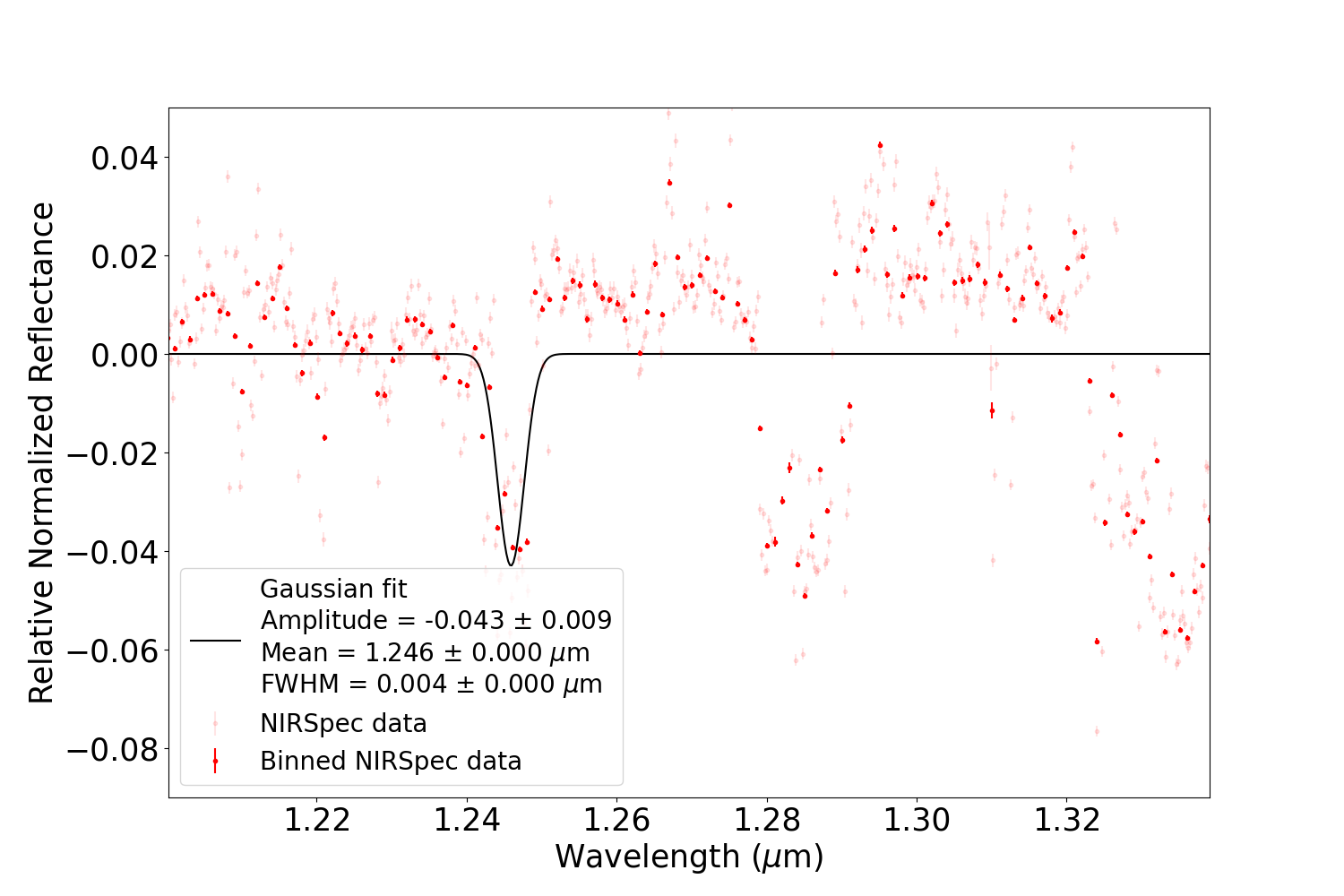}
    \caption{Relative normalized reflectance spectrum of Psyche near 1.25-$\mu$m. Binned spectra are binned to 1 nm.  }
    \label{fig:SFig1}
\end{figure}
\FloatBarrier

\FloatBarrier
\begin{figure}[!h]
    \centering
    \includegraphics[width=.5\linewidth]{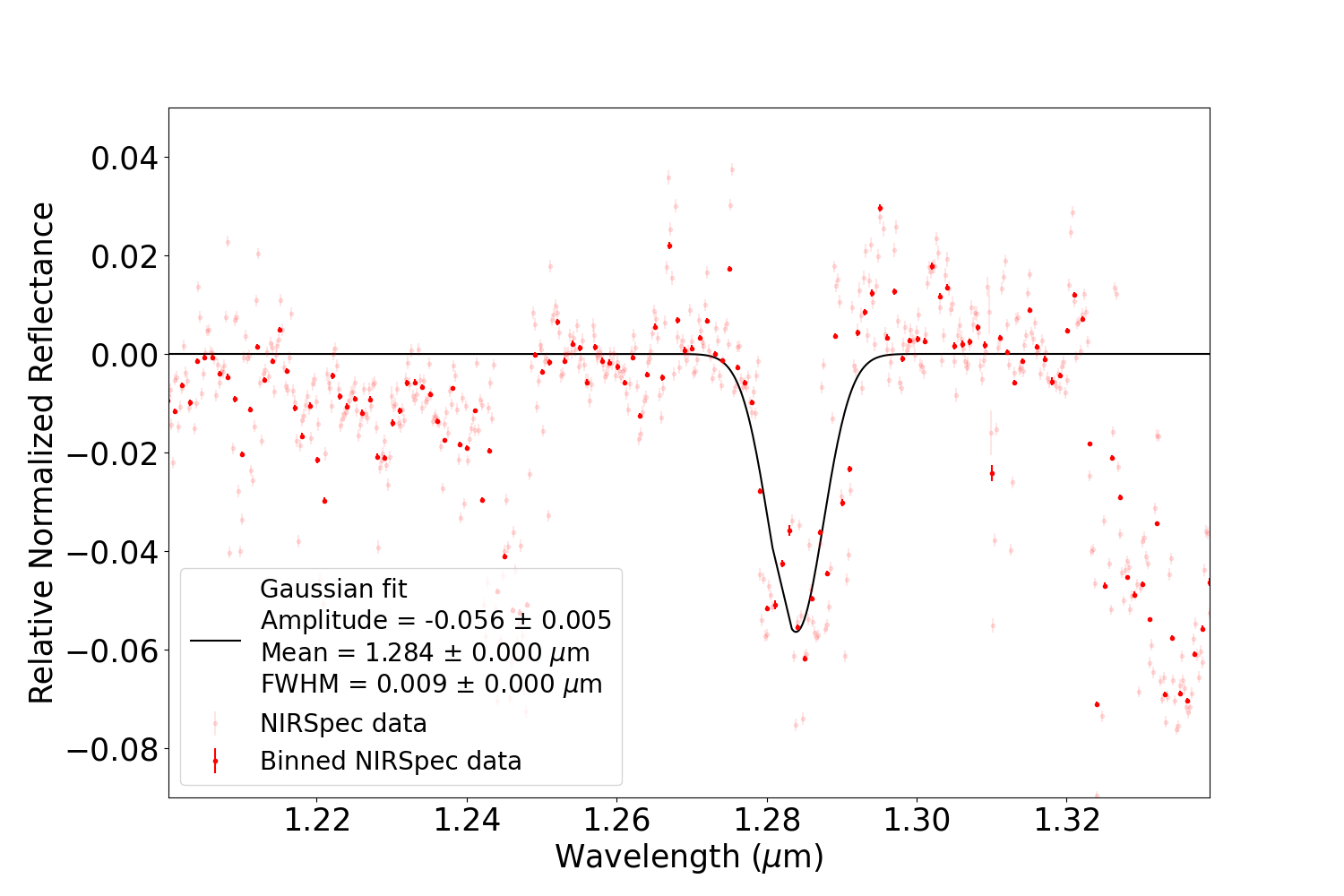}
    \caption{Relative normalized reflectance spectrum of Psyche near 1.28-$\mu$m. Binned spectra are binned to 1 nm.  }
    \label{fig:FigS2}
\end{figure}
\FloatBarrier

\FloatBarrier
\begin{figure}[!h]
    \centering
    \includegraphics[width=.5\linewidth]{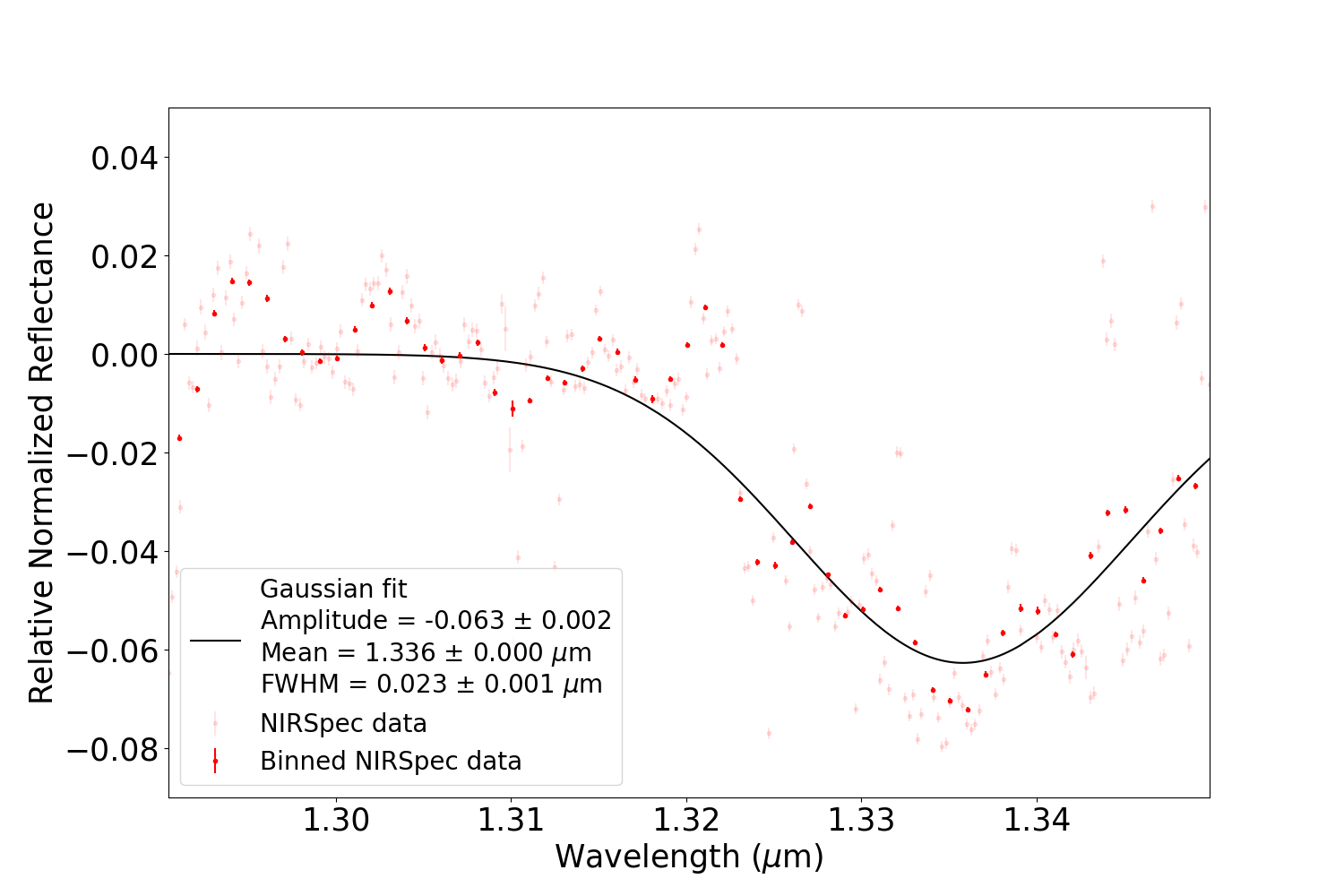}
    \caption{Relative normalized reflectance spectrum of Psyche near 1.34-$\mu$m. Binned spectra are binned to 1 nm.  }
    \label{fig:FigS3}
\end{figure}
\FloatBarrier

\FloatBarrier
\begin{figure}[!h]
    \centering
    \includegraphics[width=.5\linewidth]{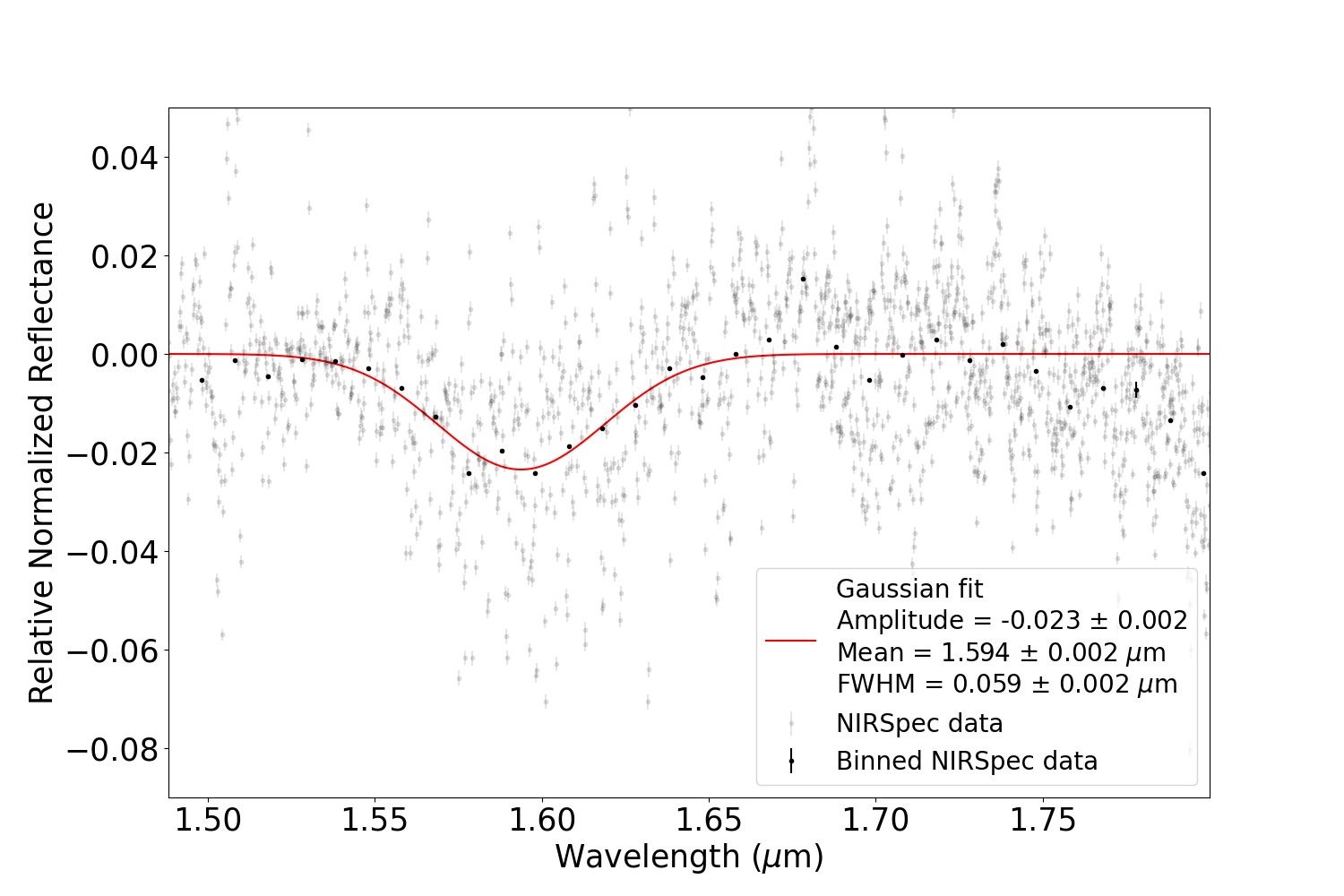}
    \caption{Relative normalized reflectance spectrum of Psyche near 1.59-$\mu$m. Binned spectra are binned to 10 nm.  }
    \label{fig:FigS4}
\end{figure}
\FloatBarrier

\FloatBarrier
\begin{figure}[!h]
    \centering
    \includegraphics[width=.5\linewidth]{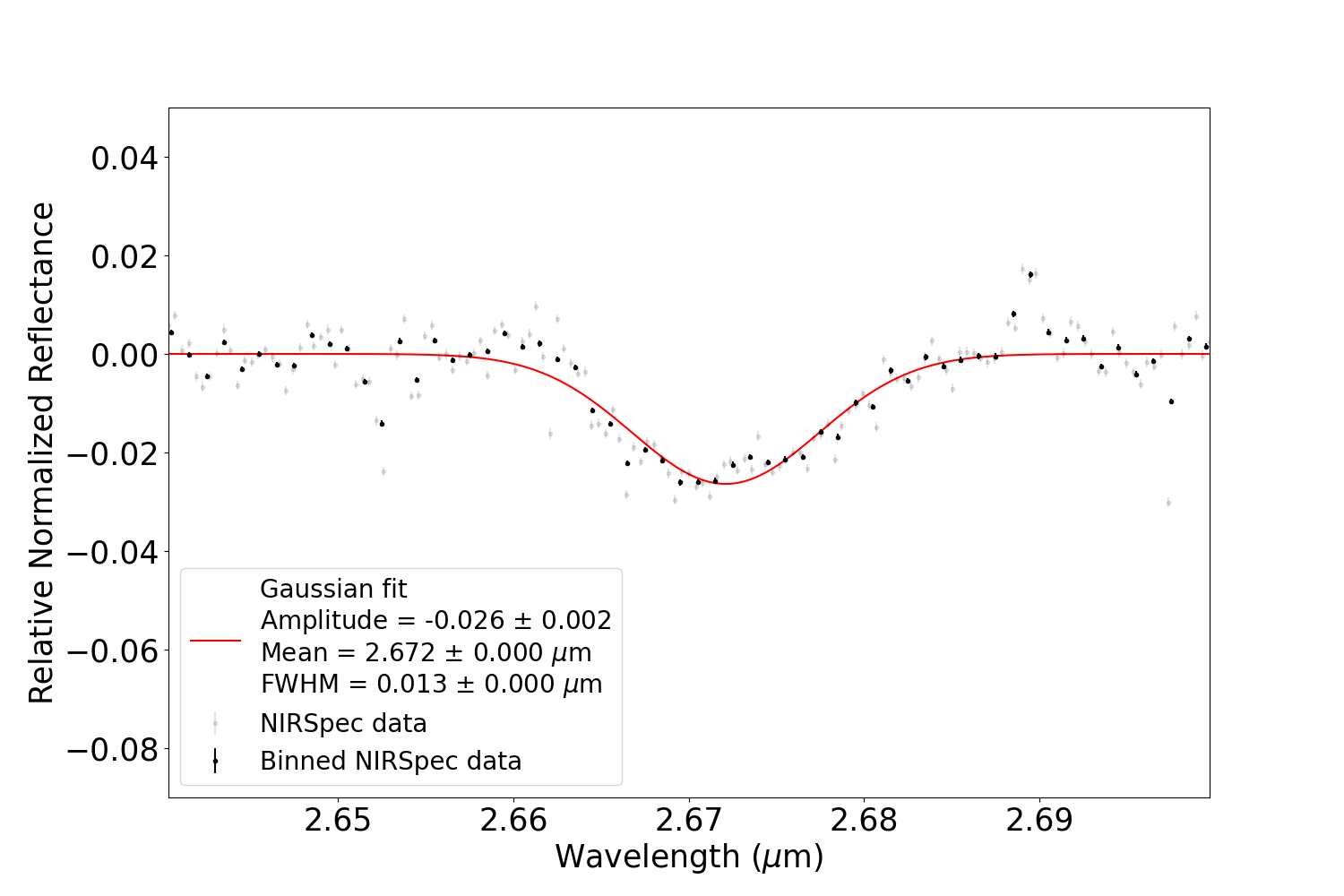}
    \includegraphics[width=.5\linewidth]{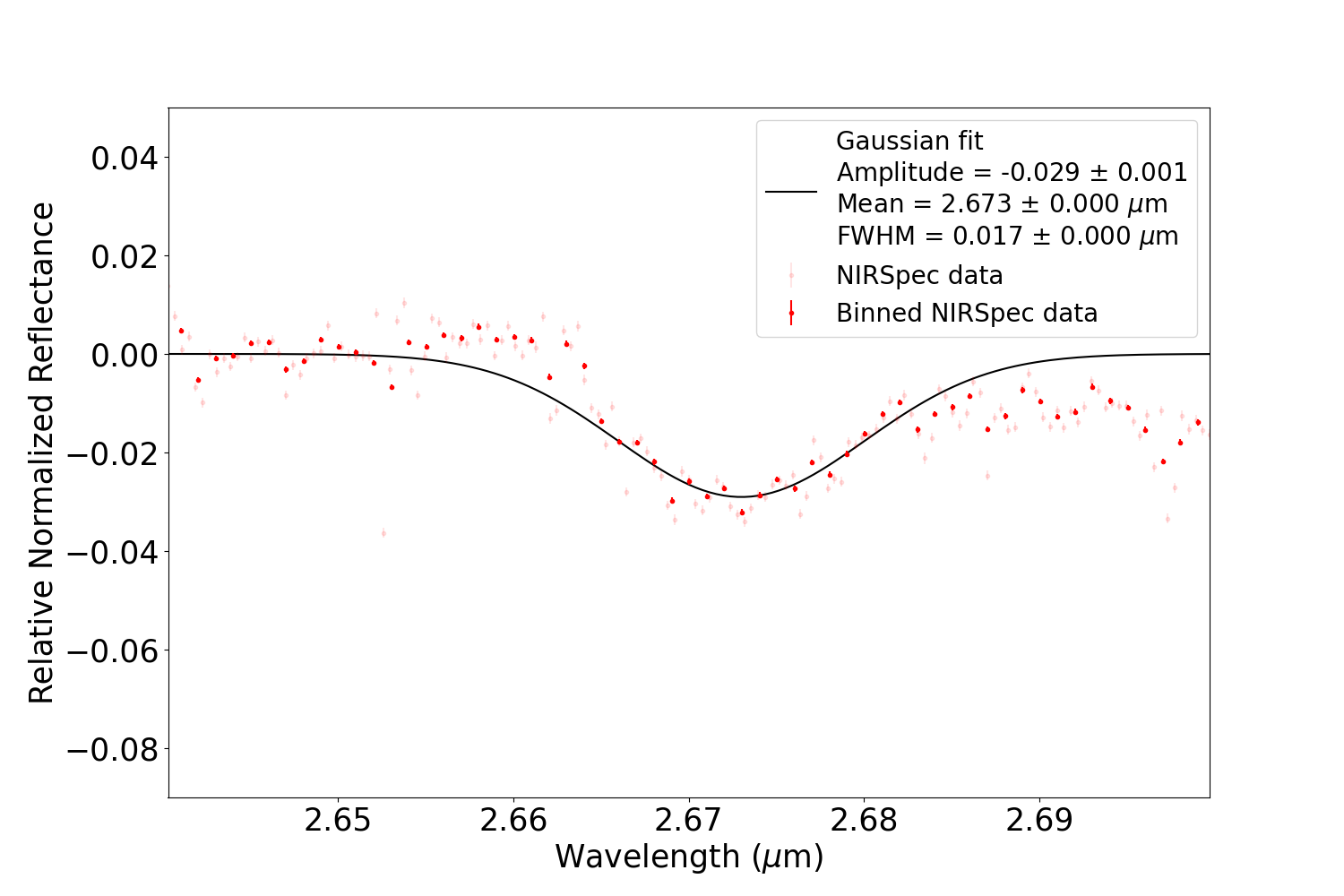}
    \caption{Relative normalized reflectance spectrum of Psyche near 2.67-$\mu$m. (Top) observation 1 (Bottom) observation 2. Binned spectra are binned to 1 nm.  }
    \label{fig:FigS6}
\end{figure}
\FloatBarrier

\FloatBarrier
\begin{figure}[!h]
    \centering
    \includegraphics[width=.5\linewidth]{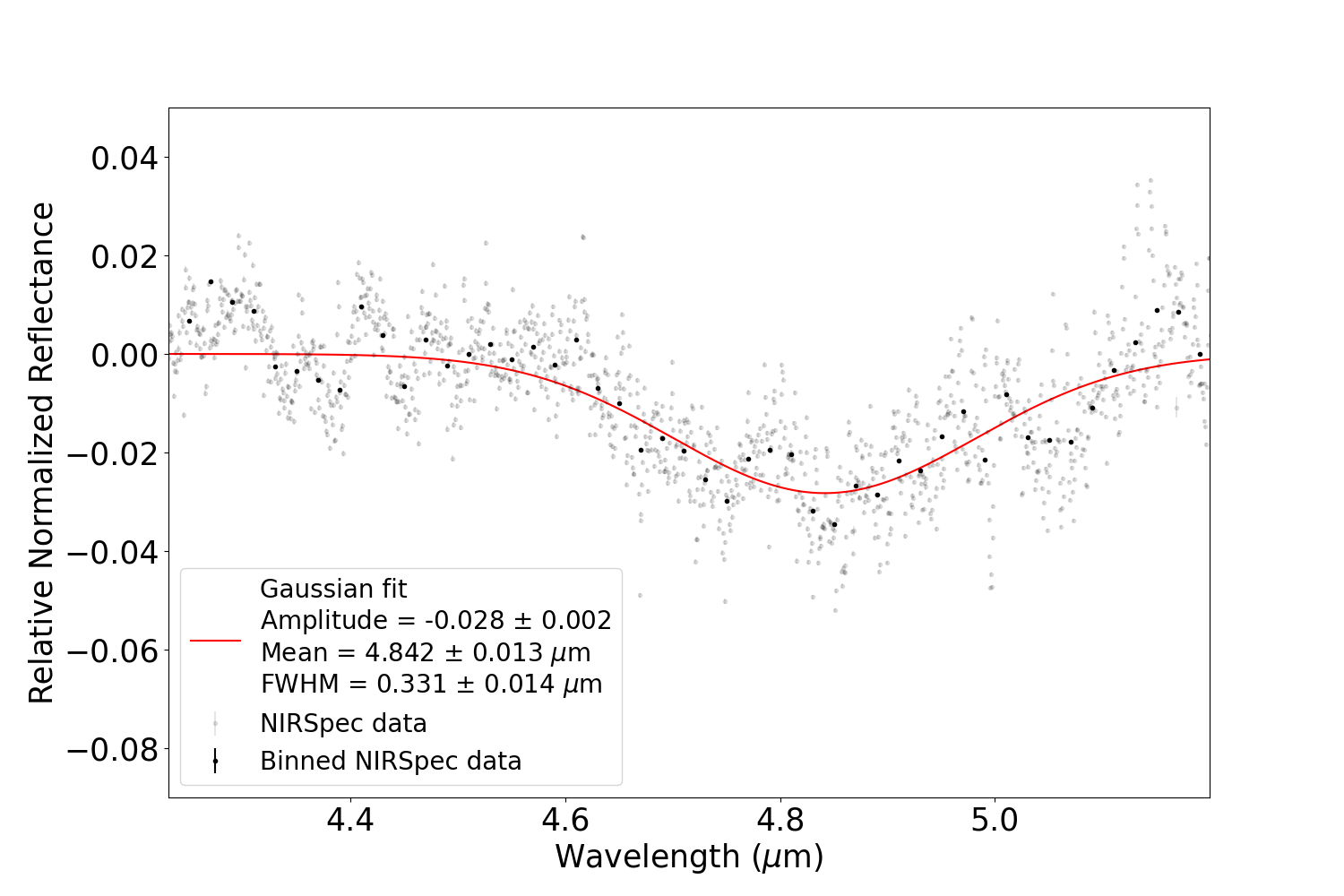}
    \includegraphics[width=.5\linewidth]{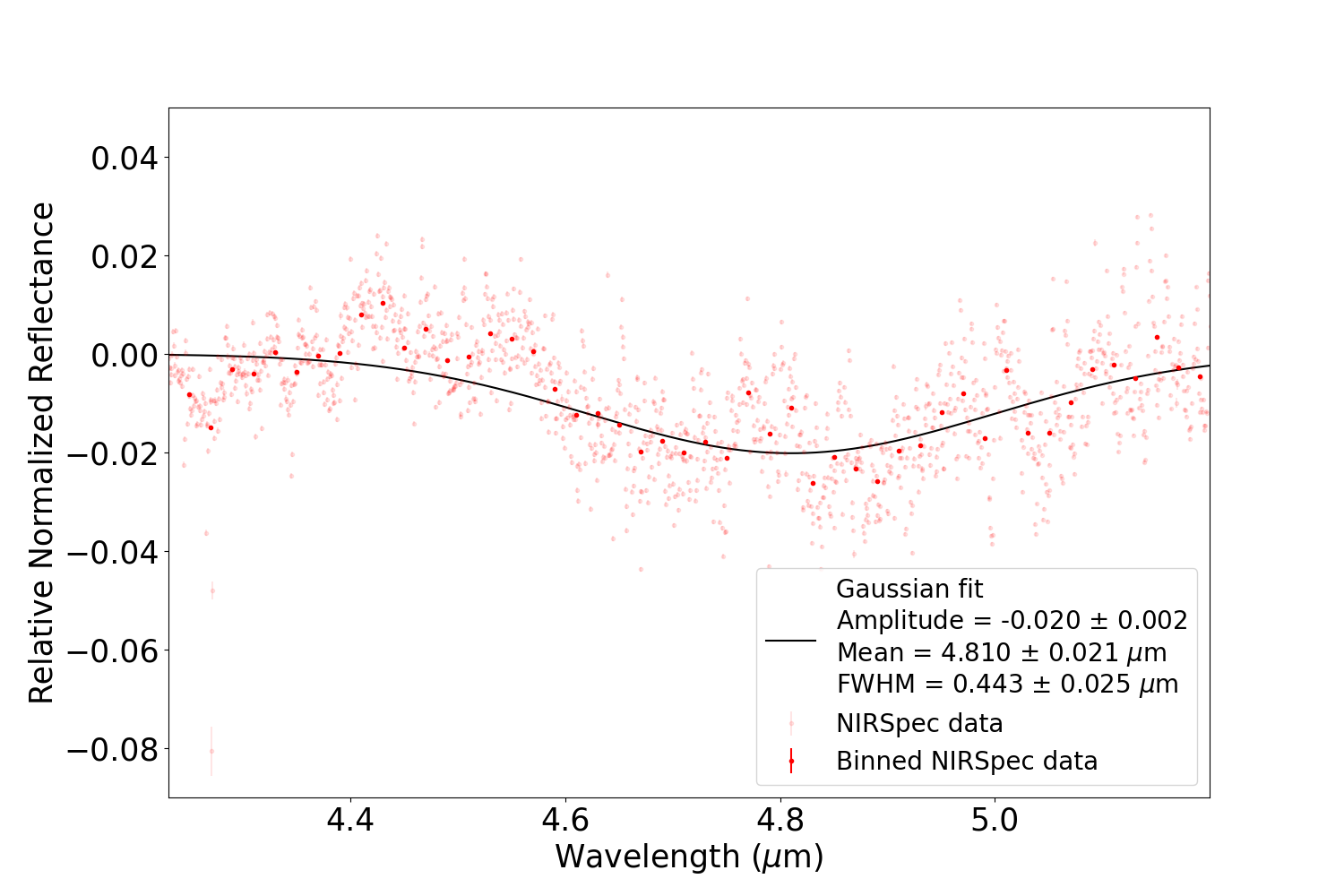}
    \caption{Relative normalized reflectance spectrum of Psyche near 4.8-$\mu$m. (Top) observation 1 (Bottom) observation 2. Binned spectra are binned to 20 nm.  }
    \label{fig:FigS8}
\end{figure}
\FloatBarrier

\FloatBarrier
\begin{figure}[!h]
    \centering
    \includegraphics[width=.5\linewidth]{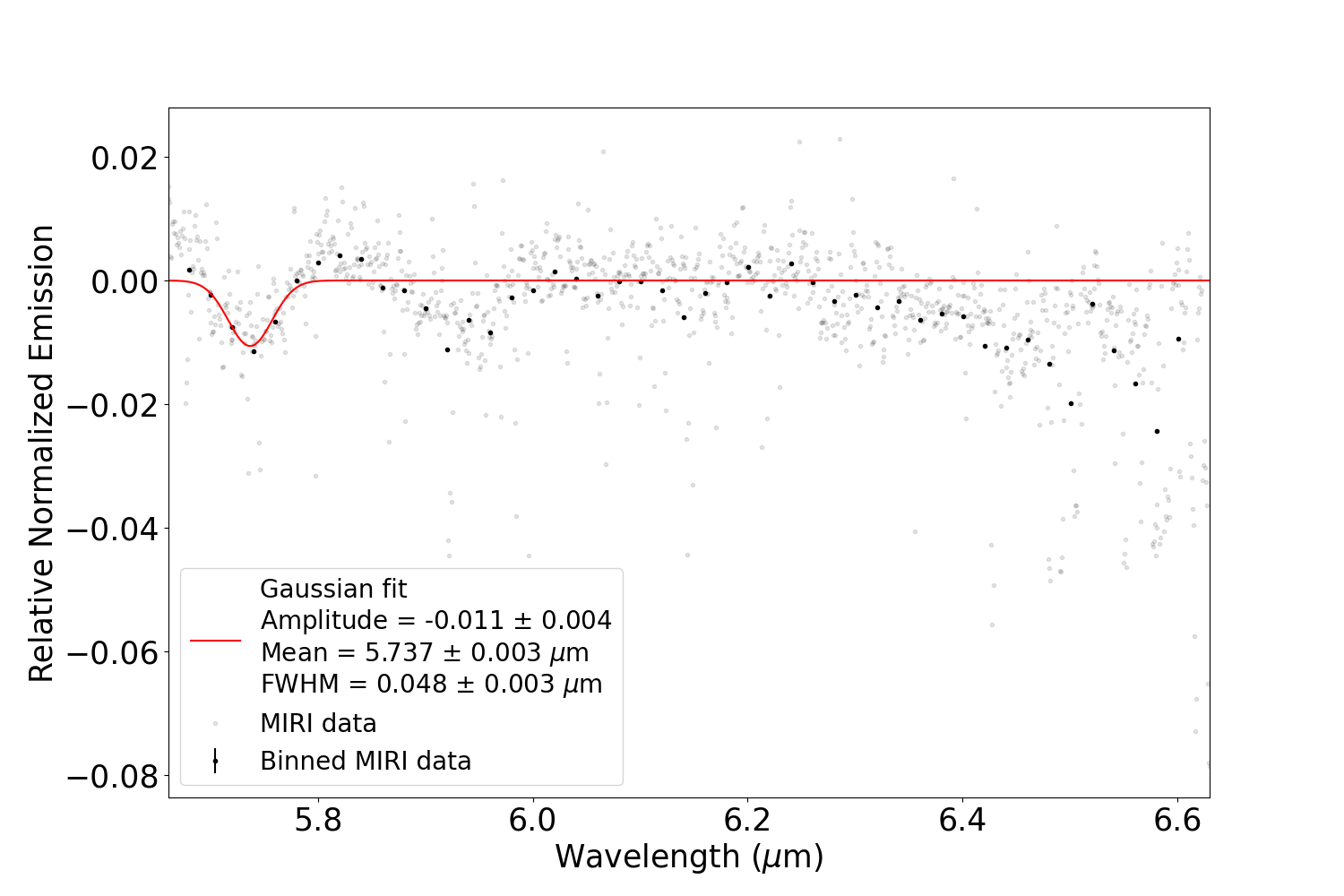}
    \includegraphics[width=.5\linewidth]{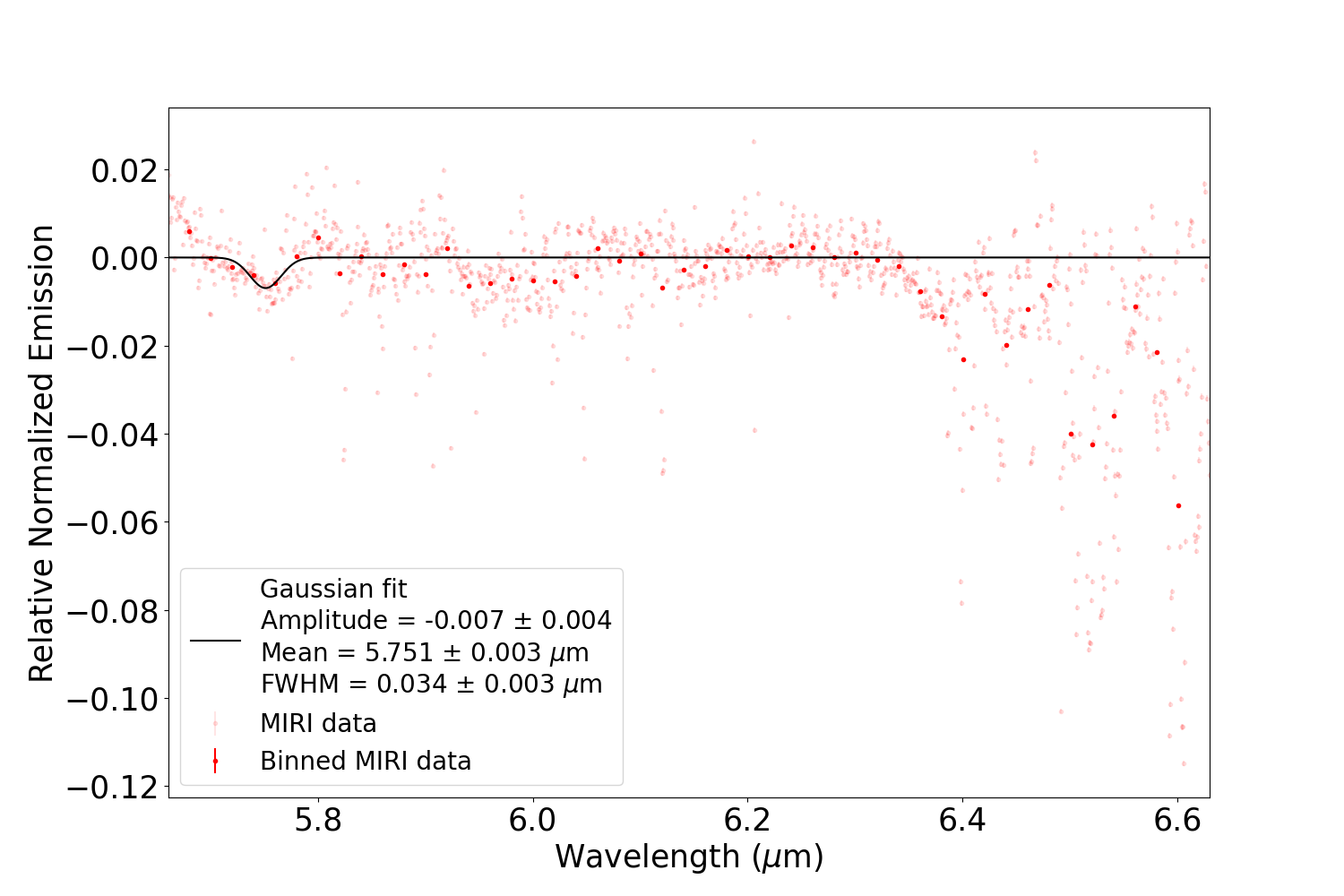}
    \caption{Relative normalized emission spectrum of asteroid (16) Psyche produced by dividing the MIRI channel 1 flux by a 2nd-order polynomial masking the region 5.9 - 6.1 $\mu$m and beyond 6.33 $\mu$m for (Top) observation 1 (Bottom) observation 2. The data are binned to a resolution of 20 nm. }
    \label{fig:FigS9}
\end{figure}
\FloatBarrier

\FloatBarrier
\begin{figure}[!h]
    \centering
    \includegraphics[width=.5\linewidth]{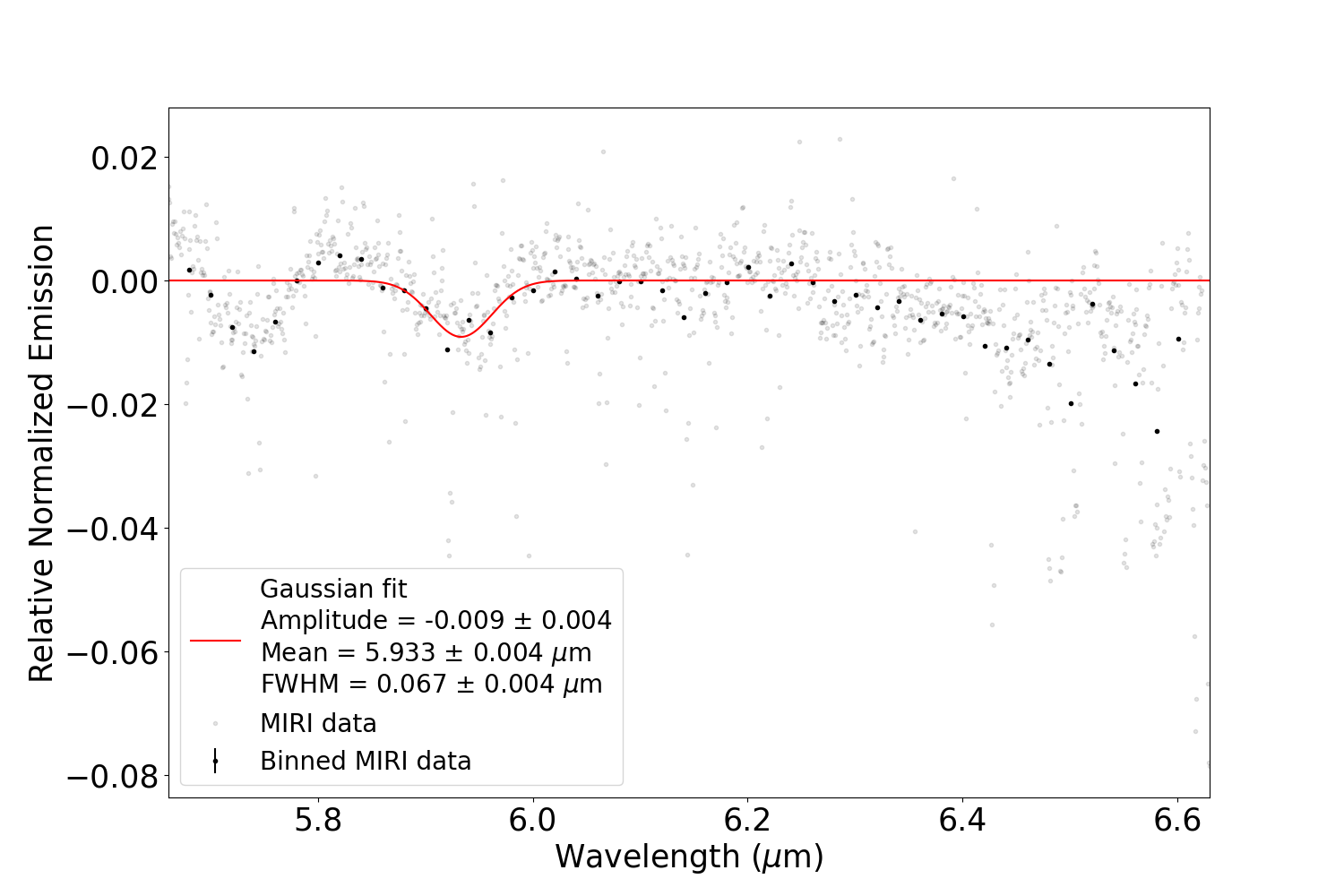}
    \includegraphics[width=.5\linewidth]{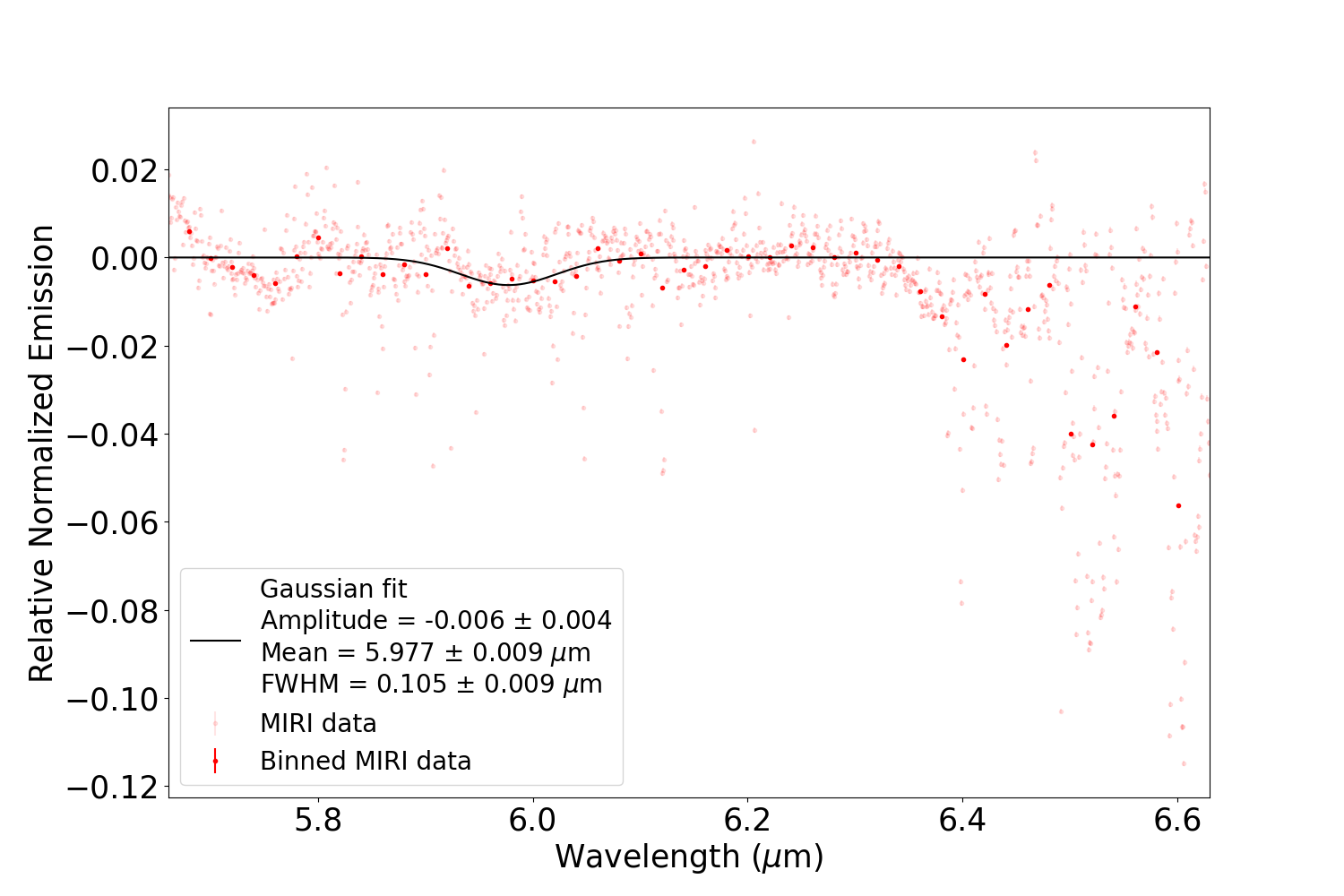}
\textbf{    \caption{Relative normalized emission spectrum of asteroid (16) Psyche produced by dividing the MIRI channel 1 flux by a 2nd-order polynomial masking the region 5.9 - 6.1 $\mu$m and beyond 6.33 $\mu$m for (Top) observation 1 (Bottom) observation 2. The data are binned to a resolution of 20 nm. }
    \label{fig:FigS9}}
\end{figure}
\FloatBarrier

\end{document}